\documentclass[journal]{IEEEtran}
\usepackage[utf8]{inputenc}
\usepackage{graphicx}
\usepackage{amsmath}
\usepackage{algorithm}
\usepackage{cancel}
\usepackage[hidelinks]{hyperref}
\usepackage{soul}
\usepackage[noend]{algpseudocode}
\usepackage{xcolor}
\usepackage{babel} 
\usepackage{multirow}
\usepackage{subcaption}
\usepackage{tabularx}
\usepackage{booktabs}
\usepackage{cite}
\usepackage{url}
\makeatletter
\def\BState{\State\hskip-\ALG@thistlm}
\makeatother

\usepackage{acro}
\DeclareAcronym{4p}{
    short = 4P,
    long = public-private people partnership, 
}

\DeclareAcronym{ac-rlnc}{
    short = AC-RLNC,
    long = Adaptive and Casual Network Coding ,
}

\DeclareAcronym{ap}{
    short = AP,
    long = access point ,
}

\DeclareAcronym{atcll}{
    short = ATCLL,
    long = Aveiro Tech City Living Lab ,
}

\DeclareAcronym{bss}{
    short = BSS ,
    long = basic service set,
}   

\DeclareAcronym{cam}{
    short = CAM ,
    long = cooperative awareness message,
}

\DeclareAcronym{cv2x}{
    short = C-V2X,
    long = cellular-vehicle-to-everything ,
}

\DeclareAcronym{dbms}{
    short = DBMS,
    long = database management system ,
}

\DeclareAcronym{dcu}{
    short = DCU ,
    long = data-collection unit,
}

\DeclareAcronym{denm}{
    short = DENM,
    long = decentralized environmental notification message ,
}

\DeclareAcronym{ev}{
    short = EV,
    long = emergency vehicle,
}

\DeclareAcronym{iab}{
    short = IAB,
    long = Integrated and Access Backhaul,
}

\DeclareAcronym{icn}{
    short = ICN,
    long = information-centric networking,
}

\DeclareAcronym{iot}{
    short = IoT ,
    long = Internet-of-things,
}

\DeclareAcronym{iiot}{
    short = IIoT ,
    long = industrial Internet-of-things,
}

\DeclareAcronym{its-s}{
    short = ITS-S ,
    long = Intelligent Transport System Station,
}

\DeclareAcronym{lifo}{
    short = LIFO, 
    long = last in first out,
}

\DeclareAcronym{llc}{
    short = LLC, 
    long = Low Latency Communications,
}

\DeclareAcronym{los}{
    short = LoS, 
    long = line-of-sight,
}

\DeclareAcronym{mec}{
    short = MEC ,
    long = multi-access edge computing,
}

\DeclareAcronym{mqtt}{
    short = MQTT,
    long = message queue telemetry transport,
}

\DeclareAcronym{ndn}{
    short = NDN,
    long = named data network-based,
}

\DeclareAcronym{nsa}{
    short = NSA ,
    long = non-standalone,
}

\DeclareAcronym{obu}{
    short = OBU,
    long = on-board unit,
}

\DeclareAcronym{ocb}{
    short = OCB,
    long = outside the context of a BSS,
}

\DeclareAcronym{otaa}{
    short = OTAA,
    long = over-the-air activation,
}

\DeclareAcronym{ovs}{
    short = OVS ,
    long = Open vSwitch ,
}

\DeclareAcronym{p2i}{
    short = P2I,
    long = people-to-infrastructure,
}

\DeclareAcronym{p2p}{
    short = P2P,
    long = people-to-people,
}

\DeclareAcronym{pdr}{
    short = PDR,
    long = packet delivery ratio ,
}

\DeclareAcronym{pgp}{
    short = PGP ,
    long = Pretty Good Privacy ,
}

\DeclareAcronym{poa}{
    short = PoA ,
    long = point of access,
}

\DeclareAcronym{ran}{
    short = RAN ,
    long = radio access network,
}

\DeclareAcronym{rf}{
    short = RF,
    long = radio frequency,
}

\DeclareAcronym{rssi}{
    short = RSSI,
    long = received signal strength indication,
}

\DeclareAcronym{rsu}{
    short = RSU,
    long = road-side unit,
}

\DeclareAcronym{rtsp}{
    short = RTSP,
    long = real-time streaming protocol,
}

\DeclareAcronym{ru}{
    short = RU,
    long = radio unit,
}

\DeclareAcronym{sa}{
    short = SA ,
    long = standalone,
}

\DeclareAcronym{sbc}{
    short = SBC ,
    long = single-board computer,
}

\DeclareAcronym{sdn}{
    short = SDN,
    long = software-defined network,
}

\DeclareAcronym{sdr}{
    short = SDR ,
    long = software-defined radio ,
}

\DeclareAcronym{smf}{
    short = SMF ,
    long = single-mode optical fiber,
}

\DeclareAcronym{tcp}{
    short = TCP ,
    long = transmission control protocol,
}

\DeclareAcronym{tsn}{
    short = TSN ,
    long = time-sensitive network, 
}

\DeclareAcronym{uav}{
    short = UAV,
    long = unmanned aerial vehicle,
}

\DeclareAcronym{ue}{
    short = UE,
    long = user equipment ,
    long-plural-form = user equipment ,
}

\DeclareAcronym{urllc}{
    short = URLLC,
    long = Ultra Reliable LLC ,
}

\DeclareAcronym{v2i}{
    short = V2I ,
    long = vehicle-to-infrastructure,
}

\DeclareAcronym{v2v}{
    short = V2V ,
    long = vehicle-to-vehicle,
}

\DeclareAcronym{v2x}{
    short = V2X,
    long = vehicle-to-everything ,
}

\DeclareAcronym{vam}{
    short = VAM,
    long = VRU awareness message,
}

\DeclareAcronym{vanet}{
    short = VANET ,
    long = vehicular ad-hoc network,
}

\DeclareAcronym{vcm}{
    short = VCM,
    long = VANET connection manager ,
}

\DeclareAcronym{vm}{
    short = VM,
    long = virtual machine,
}

\DeclareAcronym{vru}{
    short = VRU,
    long = vulnerable road user ,
}

\usepackage[percent]{overpic}
\definecolor{figgreen}{RGB}{151, 208, 119}
\definecolor{figred}{RGB}{218, 0, 13}

\DeclareGraphicsExtensions{.pdf,.png,.jpg}
\graphicspath{{./figures/}}

\title{Aveiro Tech City Living Lab: A Communication, Sensing and Computing Platform for City Environments}

\author{\IEEEauthorblockN{
Pedro Rito\IEEEauthorrefmark{2},
Ana Almeida\IEEEauthorrefmark{1}\IEEEauthorrefmark{2},
Andreia Figueiredo\IEEEauthorrefmark{1}\IEEEauthorrefmark{2},
Christian Gomes\IEEEauthorrefmark{2},
Pedro Teixeira\IEEEauthorrefmark{1}\IEEEauthorrefmark{2},
Rodrigo Rosmaninho\IEEEauthorrefmark{1}\IEEEauthorrefmark{2},
Rui Lopes\IEEEauthorrefmark{1}\IEEEauthorrefmark{2},
Duarte Dias\IEEEauthorrefmark{2},
Gonçalo Vítor\IEEEauthorrefmark{1}\IEEEauthorrefmark{2},
Gonçalo Perna\IEEEauthorrefmark{1}\IEEEauthorrefmark{2},
Miguel Silva\IEEEauthorrefmark{2},
Carlos Senna\IEEEauthorrefmark{2},
Duarte Raposo\IEEEauthorrefmark{2},
Miguel Luís\IEEEauthorrefmark{2}\IEEEauthorrefmark{3},
Susana Sargento\IEEEauthorrefmark{1}\IEEEauthorrefmark{2}
Arnaldo Oliveira\IEEEauthorrefmark{1}\IEEEauthorrefmark{2}, 
Nuno Borges de Carvalho\IEEEauthorrefmark{1}\IEEEauthorrefmark{2}
}

\IEEEauthorblockA{\IEEEauthorrefmark{1}{Departamento de Eletr\'onica, Telecomunica\c{c}\~{o}es e Inform\'atica, Universidade de Aveiro, 3810-193 Aveiro, Portugal}}

\IEEEauthorblockA{\IEEEauthorrefmark{2}{Instituto de Telecomunica\c{c}\~oes, 3810-193 Aveiro, Portugal}}

\IEEEauthorblockA{\IEEEauthorrefmark{3}{ISEL - Instituto Superior de Engenharia de Lisboa, Instituto Politécnico de Lisboa, 1959-007 Lisboa, Portugal}}
}

\date{May 2022}

\begin{document}

\maketitle

\begin{abstract}

This article presents the deployment and experimentation architecture of the Aveiro Tech City Living Lab (ATCLL) in Aveiro, Portugal. 
This platform comprises a large number of \acl{iot} devices with communication, sensing and computing capabilities. The communication infrastructure, built on fiber and Millimeter-wave (mmWave) links, integrates a communication network with radio terminals (WiFi, ITS-G5, C-V2X, 5G and LoRa(WAN)), multiprotocol, spread throughout 44 connected points of access in the city. Additionally, public transportation has also been equipped with communication and sensing units. All these points combine and interconnect a set of sensors, such as mobility (Radars, Lidars, video cameras) and environmental sensors. Combining edge computing and cloud management to deploy the services and manage the platform, and a data platform to gather and process the data, the living lab supports a wide range of services and applications: \acl{iot}, intelligent transportation systems and assisted driving, environmental monitoring, emergency and safety, among others. This article describes the architecture, implementation and deployment to make the overall platform to work and integrate researchers and citizens. Moreover, it showcases some examples of the performance metrics achieved in the city infrastructure, the data that can be collected, visualized and used to build services and applications to the cities, and, finally, different use cases in the mobility and safety scenarios.

\end{abstract}


\section{Introduction}
The concept of smart cities is not new~\cite{bib:Harrison10}. In the last decade several cities have gone through some form of digitisation and sensing to analyse their behavior and the one of their citizens, and be able to actuate to processes that are not optimized, such as in the environment, health and mobility areas. A key pillar in this process is the interconnection of the several city elements.

Over the years there have been numerous research contributions related to connectivity in different scenarios, including mobile scenarios with people, bicycles, and vehicles, but very few of them offer actual experimental results to support their claims, even if at a small scale. The large majority of existing works still rely on numerical computations and computer simulations, not addressing the impairments of real environments. 

The level of research that can be done in real environments can leverage the interaction with citizens and their services usability. On one side, research can be tested in real environments with real users, making them a beneficial part of the research. On another side, the applicability of the research can be tested and improved while it is being deployed and assessed in real environments. 

In a urban living lab, multiple stakeholders (citizens, researchers, business, authorities, and city managers) form \acp{4p} to solve problems, collaborate, cooperate, and innovate in a real-life context~\cite{link_liv_lab}. Urban living labs emerge with the increasing demand to solve urban issues, innovate, and make cities more inclusive. This is the purpose of our work: we have been deploying an advanced, large-scale communication, sensing and computation infrastructure, spread throughout the city of Aveiro in Portugal, that will be at the service of researchers, digital industries, startups, scale-ups, R\&D centers, entrepreneurs and other stakeholders interested in developing, testing or demonstrating concepts, products or services, to solve the city-related problems.

This article presents the current deployment of the Aveiro Tech City Living Lab (ATCLL)\footnote{\url{https://aveiro-living-lab.it.pt/}}. It is supported by state-of-the-art fiber link technology (spread across 16~km in the city), reconfigurable radio units, 5G-NR radio, and 5G network services. The access infrastructure covers 44~strategic points in the urban area of Aveiro, in the form of Smart Lamp Posts or wall boxes on building facades with communication technologies, edge-based computing units and sensors. The communication infrastructure integrates a communication network with radio terminals, multiprotocol (5G, 4G, ITS-G5, WiFi, and LoRaWAN/LoRa), spread throughout the city, connected by fiber optics to a data processing center. Buses and garbage collection vehicles have also been equipped with communication units with WiFi, ITS-G5 and 5G technologies, and sensors, which currently record mobility and environmental data, making a complete live map of these parameters in the city, and providing the required data for traffic monitoring and safe driving systems. All these points combine and interconnect a set of sensors, such as mobility sensors (GPS, traffic radars, LiDARs, and video cameras) and environmental sensors (such as temperature, humidity, pollution) with remote data collection units throughout the city, providing enough data to support a wide range of services and applications: from \ac{iot} and Internet access to citizens, to mobility and assisted driving, intelligent transportation systems, environmental monitoring, emergency and safety, among others.

This article describes the architecture, the technologies and the mechanisms developed to make the overall platform to work and integrate researchers and citizens, testing mechanisms and providing services to the city. It also describes some examples of the performance metrics achieved in the city infrastructure, the data that can be collected, visualized and used to build services and applications to the cities, and different use cases in the mobility and safety area that can be implemented and tested in the platform. 



The remainder of this paper is organized as follows. Section~\ref{sec:related_work} contains the related work on living labs and deployments throughout the world. Section~\ref{sec:arch} presents the proposed architecture of the living lab. Section~\ref{sec:communication} discusses the multi-technology communication approach with people, sensors and vehicles. This is followed by section~\ref{sec:edge} that contains the computing/edge approach, and the process to automate the deployment of services and applications. Section~\ref{sec:data_mgmt} contains the process to manage, process, and share data. Section~\ref{sec:results} discusses the obtained results and section~\ref{sec:use_cases} presents the defined use cases. Finally, section~\ref{sec:conclusions} presents our conclusions.

\section{Related Work}\label{sec:related_work}

This section presents some of the related work on living labs on smart campuses or smart cities deployed over the last years. We include a trend on the evolution of these living labs throughout this decade.

In 2010, Timo Ojala~\cite{ojala2010computing} proposed an urban and open testbed in the city of Oulu, Finland, with both WiFi and Bluetooth technologies and devices, to demonstrate the potential of such infrastructure. The main goal was to study the utilization of pervasive computing services with technology pilots and service prototypes. Through the creation of an Open Ubiquitous City Challenge, several researchers created new applications through the WiFi and Bluetooth data, contributing to this urban testbed.

The SmartSantander is a city-scale experimental research facility in the city of Santander, Spain. The SmartSantander~\cite{sanchez2014smartsantander} monitors the environment by using low-cost sensors to measure the air quality, noise level and luminosity level, and contains wireless technologies such as IEEE 802.15.4 and cellular GPRS. The project also monitors the occupancy of some outdoor parking spaces using ferromagnetic wireless sensors, and provides parking status information to the citizens using display panels or mobile applications. SmartSantander contains a smart and automated irrigation system that sustainably manages the water consumption. It also provides context-sensitive information and services using augmented reality. The SmartSantander living lab continued to grow over the years. The authors installed devices in 140 buses, taxis, and vans. With this information, they can assess environmental and traffic conditions~\cite{lanza2015large}.

In 2016 it was presented the City of Things, a smart city testbed established in Antwerp, Belgium~\cite{latre2016city}. The City of Things allowed the researchers to perform experiments on large-scale multi-technology \ac{iot} networks, data, and in the living lab. The technologies deployed are IEEE~802.1ac on 2.4 and 5~GHz, DASH7 on 433 and 868~MHz, Bluetooth (Low Energy), IEEE~802.15.4, IEEE~802.15.4g and LoRa. With this platform, they were able to monitor air quality, traffic, parking spot occupancy, and manage smart parking signs.

There are also several \ac{iot} sensing testbeds that focus more on the \ac{iot} architecture and services, with base wireless technologies such as WiFi, Bluetooth, LoRa or cellular, and environment sensing. Examples emerge from Mexico~\cite{Guadalajara2016} for environment, water and energy; from university campuses as living labs for energy management, such as in Brazil~\cite{Brazil2020}; from Antwerp~\cite{Antwerp2016} for air quality measurement, traffic monitoring through traffic counter sensors and smart parking. Dublin also contains a platform\footnote{\url{https://smartdublin.ie/}} on a business perspective, that addresses several areas such as environment and mobility.

The platform in~\cite{harbornet2014} presents a real-world testbed for research and development in vehicular networking that has been deployed successfully in the seaport of Leixões, Portugal. The testbed allows cloud-based code deployment and, through ITS-G5 technology, it allows remote network control and distributed data collection from moving container trucks, cranes, tow boats, patrol vessels, and roadside units, thereby enabling a wide range of experiments and performance analyses.

PortoLivingLab~\cite{santos2018porto} is a smart city testbed located in the city of Porto, Portugal. Through the deployment of several sensors installed in city buses and other static locations, it can collect, through ITS-G5 or LTE, data about the environment (noise and air pollution), weather, public transport, and people's flows. This originated three monitoring platforms with sensing capabilities and a common backend infrastructure. The first monitoring platform is a crowdsensing research tool that collects data from participants’ smartphones through WiFi and cellular. The second one allows monitoring the environment through WiFi. The last one contains a network of buses, each one equipped with an \ac{obu} in order to achieve \ac{v2v} and \ac{v2i} communication.

PASMO~\cite{Pasmo2017} is a highway ITS road between the seaside and Aveiro/Ílhavo, in Portugal, providing vehicular communications through ITS-G5. This road enables the test of vehicular communications and road automotive use cases. The living lab presented in this paper, in the city of Aveiro, integrates with PASMO in the highway.

On the 5G perspective, l'Aquila in Italy has been proposed as a living lab for 5G experimentation~\cite{lAquila2018}, but only the fiber optical ring has been presented, as well as use cases to be tested. Bristol in UK has also a testbed that is being used on a business perspective\footnote{\url{https://www.bristol.gov.uk/}}, which encompasses some technologies, such as 5G in specific places. 

On a different perspective, living labs for autonomous driving have also been deployed. The work in~\cite{Ljubliana2020} presents the Autonomous Vehicles — AV Living Lab that spreads across the shopping center BTC in Ljubljana, Slovenia, with autonomous vehicle demonstration events.

Our approach in this paper goes beyond these several examples in different aspects: (i)~the Aveiro TechCity Living Lab encompasses different communication technologies for people, sensors and vehicles, simultaneously; (ii)~it includes different sensors, in a fusion approach, addressing the monitoring with different perspectives and higher accuracy; (iii)~it encompasses both access, edge, and core infrastructure, and both network and data management; (iv)~it is programmable and with third-party seamless integration; and (v)~it addresses different use case scenarios, from environment to mobility and self-driving. 

\section{Aveiro Tech City Living Lab Architecture}\label{sec:arch}

\Ac{atcll} has been designed with the following three main objectives: (i)~to build a communication, sensing, and computing infrastructure connected by fiber optics integrating a communication multiprotocol network with radio terminals for short, medium and long-range communication; (ii)~to implement a sensing platform capable of understanding the environment quality and citizens’ behaviour within the city, and providing new solutions towards efficient traffic management, intelligent transportation systems and citizens’ safety; and (iii)~to provide an open platform for third-party partners to test their own protocols, mechanisms, prototypes, or explore collected data to build new services and applications.

The structure of the platform can be divided in the following manner: (i)~access and edge, which is supported on fiber technology, edge computing, high data-rate \ac{sdn} devices and radio units, either dedicated and \ac{sdr} based; (ii)~sensing devices and platform, static and mobile, collecting various forms of data, and computing edge units; (iii)~backhaul and core, which consists of the datacenter with core network devices and servers to manage the network; and (iv)~backend and data platform, which are composed of the data platform core with the broker, database and data processing.

\begin{figure}[!t]
	\centering
	\includegraphics[width=\columnwidth]{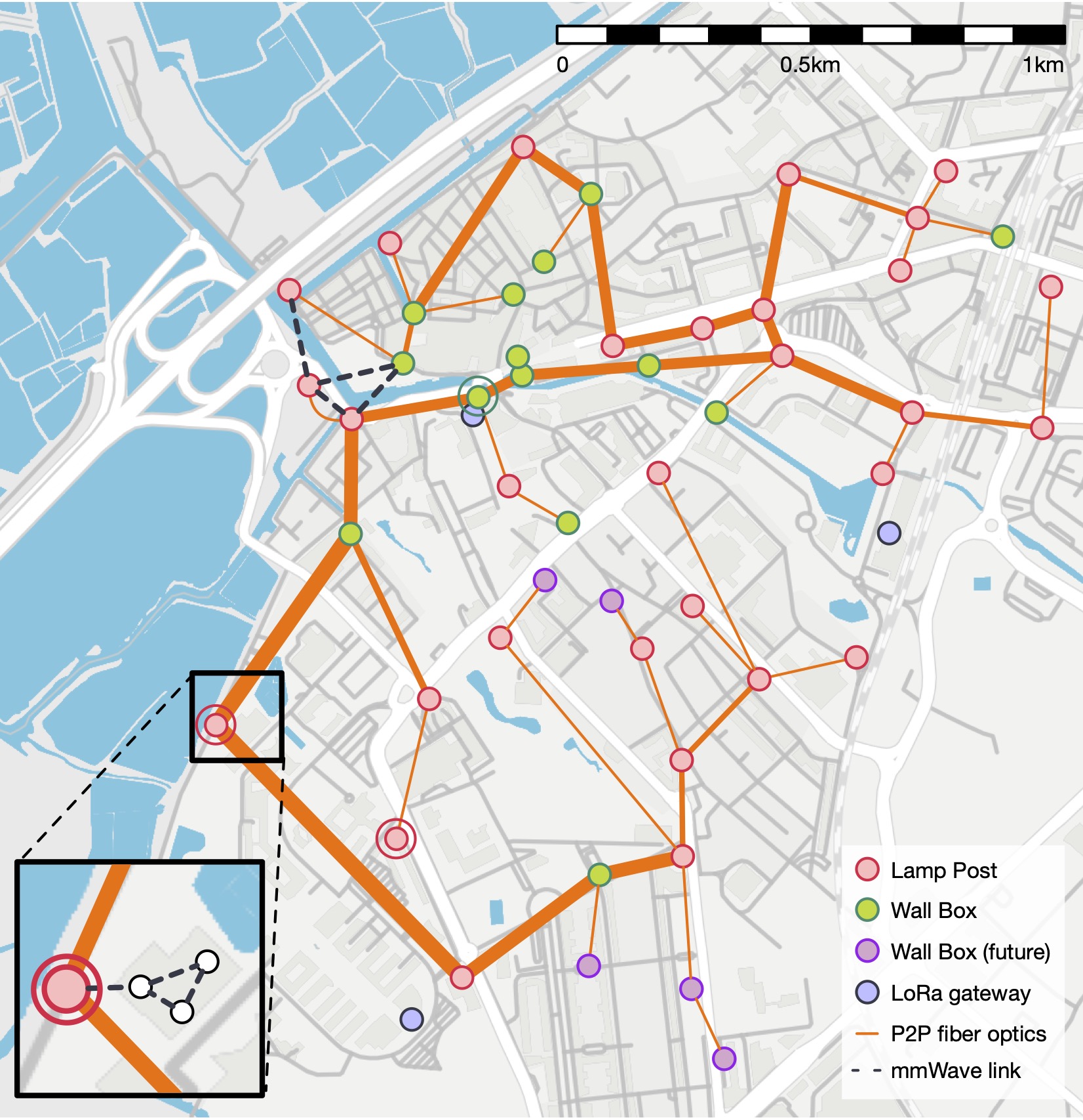} 
	\caption{ATCLL map in Aveiro. Thicker optical connections represent cables with 96 or 144 fibers, and thinner connections represent cables typically with 12 fibers.}
	\label{fig:ATCLL_map}
\end{figure}

\begin{figure}[!t]
	\centering
    \includegraphics[width=\columnwidth]{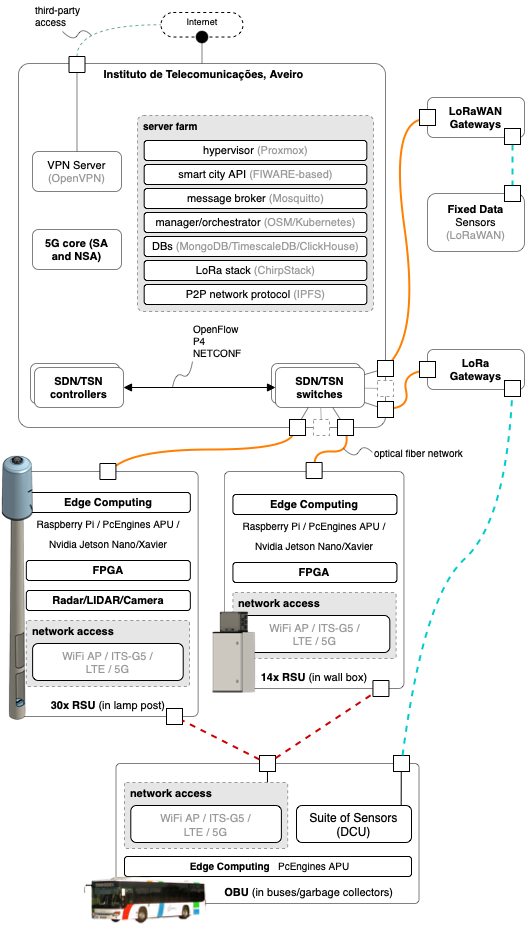} 
	\caption{ATCLL architecture.}
	\label{fig:AveiroOpenLab_architecture}
\end{figure}

Figure~\ref{fig:ATCLL_map} depicts the \ac{atcll} in the map of Aveiro, its fiber connections and the different access points placed in strategic locations in the city. Figure~\ref{fig:AveiroOpenLab_architecture} illustrates the high-level diagram of the infrastructure, architecture and services of the \ac{atcll}. The next sections describe the several components of the \ac{atcll} architecture.

\subsection{Access and Edge}
\label{sec:arch_access}

\begin{figure*}[t]
    \centering
    \includegraphics[width=\textwidth]{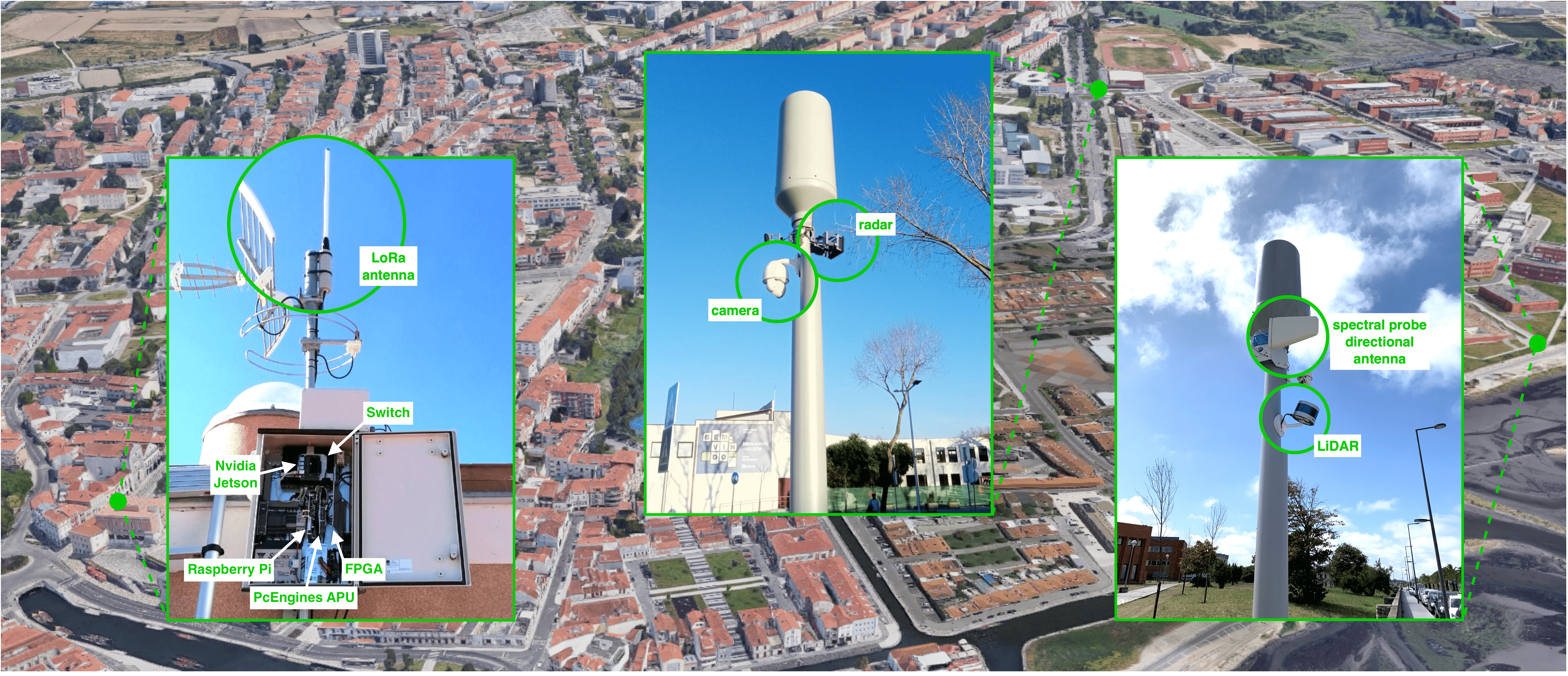}
    \caption{Wall box and Smart Lamp Post examples.}
    \label{fig:wb-slp}
\end{figure*}

The access infrastructure is based on the \ac{smf} link technology (G.652), with 16~km of length, interconnecting the edge nodes with re-configurable radio units and 5G network services, in 44 strategic places covering the urban area of Aveiro. These points are deployed in the form of Smart Lamp Posts or wall boxes on building facades. Examples of Smart Lamp Posts and wall boxes are depicted in Figure~\ref{fig:wb-slp}.

The physical infrastructure of the access network is composed of one-to-one fiber links aggregated by a switch supporting 10~GbE connections, \ac{sdn} functionality through \ac{ovs}\footnote{https://www.openvswitch.org/}, and \ac{tsn}\footnote{https://1.ieee802.org/tsn/} features configured by NETCONF. The uplink to the core is completed with 40~GbE ports. Supporting \ac{sdn} allows the control of the network by software, using open-source controllers such as Ryu\footnote{https://ryu-sdn.org/} and ONOS\footnote{https://opennetworking.org/onos/}, and control plane protocols OpenFlow and P4 Runtime\footnote{https://opennetworking.org/p4/}. In addition, network algorithms and machine learning techniques are also included to monitor and take action in the network. For instance, in the case of the mobility network, Internet connectivity and handovers in the mobile nodes are made possible by the algorithms running through the \ac{sdn} controller~\cite{MSilva2021}.

Regarding the edge points, they combine different equipment for several purposes in each Smart Lamp Post/wall box: communication access points, \ac{mec} and sensing devices. In terms of communication technologies, there are two main devices providing wireless access: \mbox{FPGA-based} \ac{sdr} units, with a wideband (300 MHz to 6 GHz) integrated \ac{rf} frontend, to provide cellular access (4G, 5G and beyond) and a PC Engines APU\footnote{https://www.pcengines.ch/apu.htm}. The PC Engines APU combines two wireless modules: WiFi and \mbox{ITS-G5}, and some with \mbox{\ac{cv2x}}). Some of the APUs also integrate 5G modules that work with a 5G experimental network in Aveiro provided by Altice MEO Operator\footnote{https://www.meo.pt/}, and some of the edge points also integrate LoRa and/or LoRaWAN gateways. Regarding \ac{mec}, every point has available a set of Nvidia Jetson Nano or Jetson Xavier with a powerful GPU for graphic-intensive processing, and a Raspberry Pi 4 aimed for the deployment of lightweight services. Finally, in terms of sensing devices there are spectral probes, traffic radars, LiDARs and video cameras.






\subsection{Sensing}
\label{sec:arch_sensing}

As referred previously, the \ac{atcll} contains static sensors in the edge nodes of the infrastructure such as spectral probes, traffic radars, LiDARs, and video cameras (as seen in Figure~\ref{fig:wb-slp}). Information coming from these devices is aggregated and correlated to give insights on the people’s flow, providing concrete elements for new solutions on public transportation, safety-critical and autonomous driving systems, and to identify problems and optimize the mobility in the city. The spectral probe is a passive radar system consisting of an antenna array comprising three dipoles forming a 360º antenna and a low-cost commercial \ac{sdr} (ADALM-PLUTO) connected to the Raspberry Pi for processing the collected data. The \mbox{FPGA-based} \ac{sdr} units can also be individually reconfigured to implement wideband, high-performance and real-time distributed spectral probes, with centralized processing of the acquired signals that can be stored and used as rich data sets for different purposes. The other sensors are described in more detail in subsection \ref{sec:edge_fusion}. A set of \acp{uav} equipped with video cameras and sensors are also considered as mobile sensing units to gather data from the city and to give support to patrolling and traffic management.

\begin{figure*}[t]
    \centering
    \includegraphics[width=\textwidth]{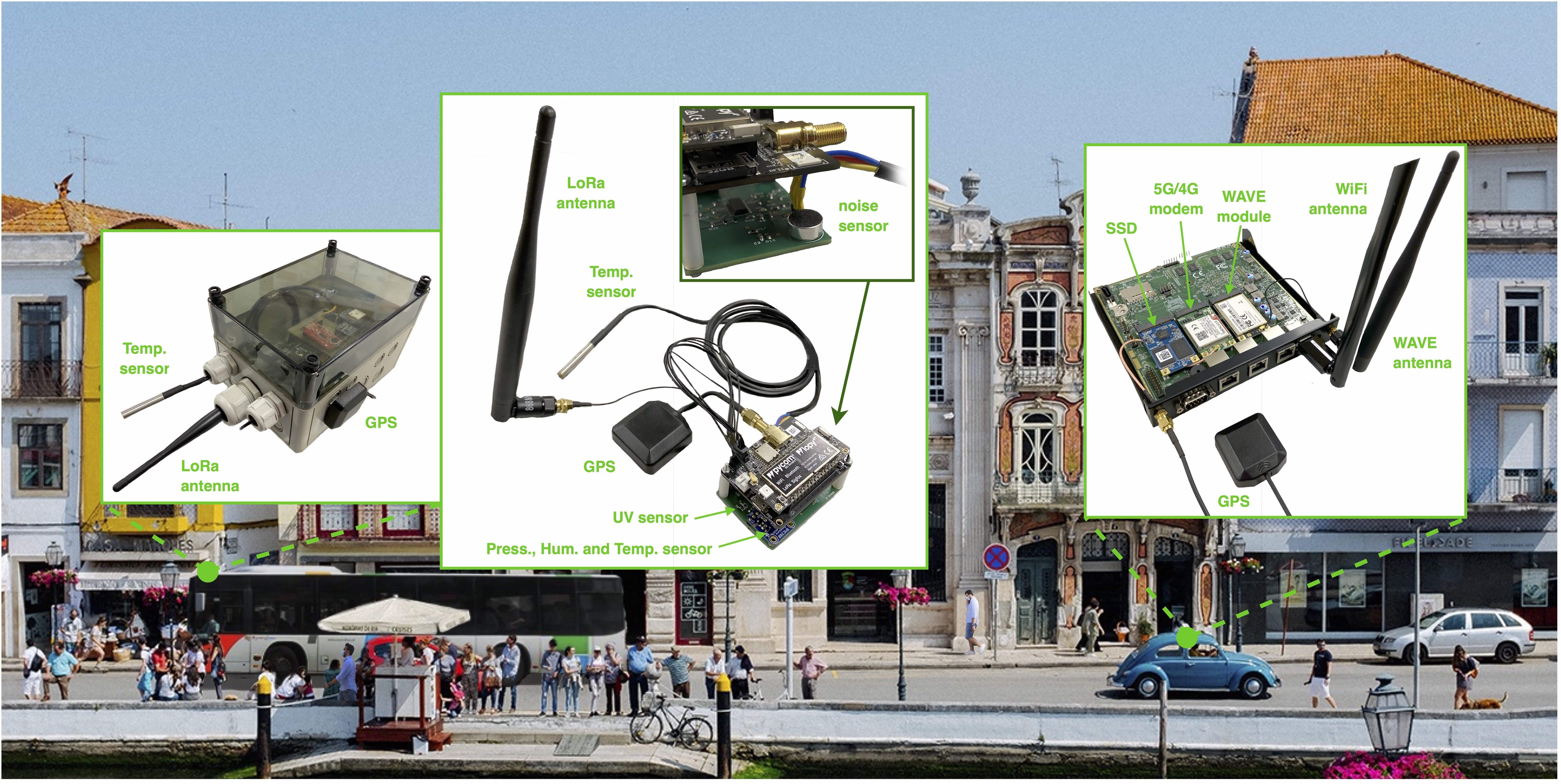}
    \caption{Data Collecting Units (DCUs) and On-Board Units (OBUs).}
    \label{fig:dcus-obus}
\end{figure*}

The \ac{atcll} also contains mobile sensing devices and other geolocation sensors installed on vehicles (buses and garbage collection vehicles) and local boats (\textit{moliceiros}). The available mobile sensing information comprises GPS location, speed and heading, temperature, humidity and air pressure, which enables the complete mobility map of the city. Communication equipment is also installed in the vehicles and local boats to transmit mobility and environment data through the static access points, making a complete live map of these parameters in the city, and providing the required data for environment sensing, traffic monitoring and safe driving systems. The mobile equipment is composed of a \ac{dcu} (Figure~\ref{fig:dcus-obus}) which integrates WiFi and LoRa communication. The vehicles also include an \acf{obu} (Figure~\ref{fig:dcus-obus}) with WiFi, ITS-G5 and cellular communication (currently some units already integrate both 4G and 5G modules) to establish the connection with the \acfp{rsu} and with other vehicles. Currently, 10 public buses are working as connected vehicles with \ac{obu}s, and an extension to 30 buses is being performed. Cars with integrated OBUs from the start are also detected and communicate with our \ac{rsu}s. The WiFi access points, both in the static access points and in mobile on board units in the vehicles, are also gathering data from people’s smartphones through their connectivity in the city.



\subsection{Backhaul and Core}
\label{sec:arch_backhaul_core}

The infrastructure is complemented with a network and data processing center located at the premises of Instituto de Telecomunicações, in Aveiro. The datacenter is composed of multiple servers managed by Proxmox VE\footnote{https://www.proxmox.com/en/proxmox-ve}, which deploy all virtual machines with the services, data platform, apps, 5G core and network functions running in the cloud, such as Kubernetes\footnote{https://kubernetes.io/}, OSM\footnote{https://osm.etsi.org/}, ChirpStack\footnote{https://www.chirpstack.io/}, among others. The \ac{sdn} controllers are also deployed in the core of the network, defining the structure of access and mobility networks, and reacting dynamically to possible changes in the demand and conditions of the network. An important feature of the infrastructure is the possibility to give access to the sensors, network devices or services to third-parties, which is allowed by an OpenVPN server.

\subsection{Backend and Data Platform}
\label{sec:arch_backed}

\begin{figure*}[t]
    \centering
    \includegraphics[width=\textwidth]{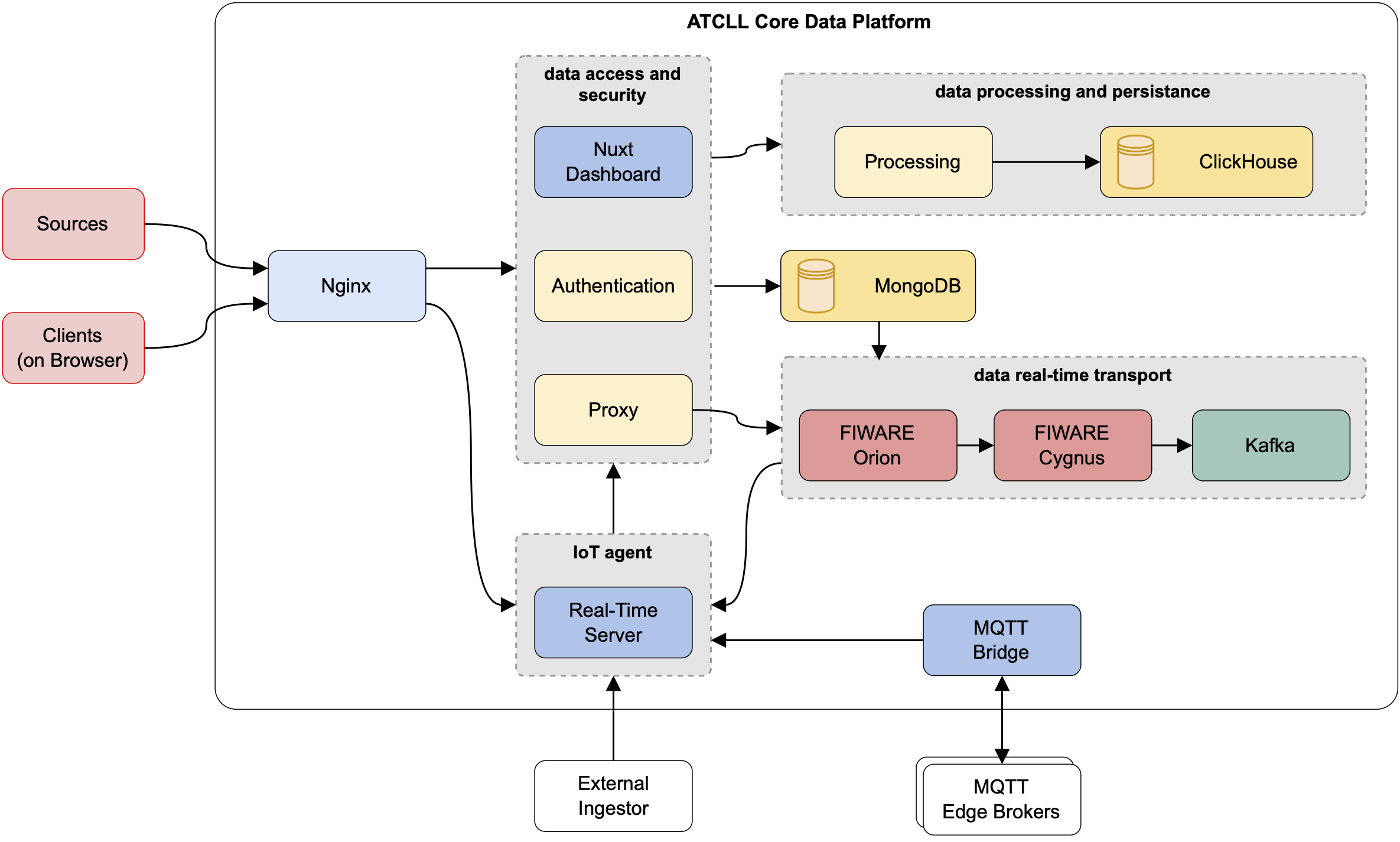} 
    \caption{ATCLL data platform diagram.}
    \label{fig:data_platform}
\end{figure*}

The management of the data collected through the ATCLL sensing devices is performed by the \emph{Core Data Platform}, as depicted in Figure~\ref{fig:data_platform}. This platform is capable of receiving, storing, and processing information from several different domains, while exposing it in a secure and efficient manner via various different interfaces. This section provides an overall overview of this platform. Further details are available in~\cite{paper_GVitor}.

As shown in Figure~\ref{fig:data_platform}, the data platform can be divided in four main modules: the data access and authentication, data persistence, data real-time transport, and the \ac{iot}-agents. The platform was designed on top of the multi-tenant model of FIWARE's Orion Context Broker\footnote{https://fiware-orion.readthedocs.io/en/master/}, the core of the real-time transport module and the point of access of the \ac{atcll} platform. This broker follows a publisher-subscriber model that allows the creation and update of entities and subscriptions, logically separated by the value present at the FIWARE-Service header of the requests, hence the \mbox{multi-tenant} model. The data platform extends this model and also uses the FIWARE-Service header to perform authentication, persistence, and access to the information separated by tenants/domains. The Orion Broker also supports hierarchical scopes, with the \texttt{FIWARE-ServicePath} header, that can be freely used with the Orion instance to divide entities and subscriptions of a FIWARE-Service by different paths. 
On the entrance of the platform, there is an Nginx\footnote{https://www.nginx.com} instance working as a reverse proxy while also managing the SSL certificates, offloading the deciphering process from the servers.

The dashboard allows the visualization of current and historical data. It is possible to visualize data collected by the sensors, static or placed in mobile elements such as buses and garbage trucks. The dashboard communicates directly with the Nginx server, forwarding the requests to the respective service. The data is obtained through the Processing server APIs and the ClickHouse\footnote{https://clickhouse.com/} database.

Furthermore, all new communications received from mobile or static sensors are registered in a centralized monitoring and alerting platform (Zabbix\footnote{\url{www.zabbix.com}} and Grafana\footnote{https://grafana.com/}), allowing the examination of the health status (time since last seen) of a given end-device and, therefore, of the system as a whole at any given time.

The data gathered from all smart city elements provides the required conditions to support services and applications, such as Internet access for all citizens, environmental sensing, mobility and emergency services, content distribution, assisted driving, among others.

\section{Communication Approach}
\label{sec:communication}

This section details the communication approach between the mobile and static elements, and how the mobile and sensing nodes are able to be seamlessly connected between themselves and to the infrastructure. 

\subsection{Communication Technologies}
\label{sec:communication_tech}

One of the main novelties of this platform is its \mbox{multi-technology} capability. It contains short-, medium- and long-range technologies that are used between their different elements. Through these radio access technologies, the infrastructure integrates sensors, people, through their mobile phones, and vehicles that support ITS-G5 technology or 5G, such as buses, bicycles in the city and \textit{moliceiros} in the Aveiro Lagoon. 

For short-range communication, the infrastructure supports ITS-G5 and WiFi. The \acp{rsu} for ITS-G5 and WiFi access points are located in all 44 city nodes represented in Figure~\ref{fig:ATCLL_map}. Additionally, 2 nodes contain also Cellular Vehicular-to-Everything (C-V2X). The \acp{rsu} and \acp{obu} are implemented using a \ac{sbc} produced by PC Engines\footnote{http://pcengines.ch}, which is equipped with an IEEE 802.11a/b/g/n mini-PCIe wireless card. This mini-PCIe wireless card is used for ITS-G5 \ac{v2x} communications. Additionally, the \ac{rsu} contains a second wireless card for the WiFi \ac{ap}, and the \ac{obu} contains an LTE CAT-1 mini-PCIe module or \mbox{5G \textit{m.2}} module and an external USB dual-band wireless adapter (Figure~\ref{fig:dcus-obus}). In order to use the wireless card compatible with a ITS-G5 standard, it requires patching the Linux kernel, to permit operating in the desired 5.9 GHz band and including the \acs{ocb}~mode support\footnote{\url{https://ctu-iig.github.io/802.11p-linux/}} (meaning \textit{\acl{ocb}}) to the \texttt{ath9k} driver. 
The LTE (4G) module is used for cellular connectivity and GPS location, since GNSS technology is included in this module. For the 5G module to work properly, the ModemManager from \textit{freedesktop.org}\footnote{https://www.freedesktop.org/wiki/Software/ModemManager/} was used. This module supports \ac{nsa} and \ac{sa} operation modes depending on additional configurations. The cellular link allows to have connectivity in the surroundings of the city and in non-covered urban areas with the \acp{rsu}' \mbox{ITS-G5} connection. 
The external wireless adapter serves as an access point for passengers inside the vehicle and for the \ac{dcu} that sends all the gathered data via WiFi to the \ac{obu}, to forward it to its final destination through the \ac{v2i} communication (ITS-G5 or 5G).

In terms of medium-range communication, apart from the deployment of the \acp{obu} (with cellular modems) as \acp{ue}, the infrastructure includes \ac{ran} deployments, where the \acp{ru} or small cells are placed inside the Smart Lamp Posts and the wall boxes on building facades. The main \ac{ran} deployment is a 5G O-RAN demonstrator with 20+ \acp{ru} built on FPGA-based \ac{sdr} boards. This distributed topology of \acp{ru} takes advantage of the fiber network linking the city nodes to the core for the fronthaul communication. Other nodes include small cells of third-party vendors. With both approaches, the \ac{atcll} offers the possibility to have private 5G network deployments for testing and development of use cases requiring cellular access. Moreover, with the integration of \ac{sdr} hardware, the infrastructure is also evolving into a testbed of beyond-5G and 6G technologies.

Finally, regarding long-range communication, the infrastructure supports LoRa communication using our custom protocol~\cite{Oliveira2017} and LoRaWAN protocol, both in the end devices (\aclp{dcu}) and in the gateways. The LoRa with custom protocol implementation hardware consists, in \acp{dcu} and gateways, of a Raspberry Pi model 3B+ \ac{sbc} with a multi-protocol radio shield that attaches a Libelium SX1272 module and an 868MHz antenna.
To maximize the range while maintaining a sensible data rate, the SX1272 module is used in mode 3, which entails a spreading factor of 10 and a coding rate of $4/5$. 


For the LoRaWAN protocol, the \acp{dcu} are based on LoPy4\footnote{https://pycom.io/product/lopy4/} (Figure~\ref{fig:dcus-obus}), and the gateways are LORIX One\footnote{https://www.lorixone.io/} or an extension to the APU board with a mPCIe card. End devices join the network via \ac{otaa} and transmit, in most cases, using a spreading factor of 9 and a coding rate of $4/5$. At the time of writing of this article, there are two LORIX One outdoor gateways placed in strategic locations in the city (Figure~\ref{fig:wb-slp}), with more gateways to be deployed in the near future.

\subsection{Sensing to Infrastructure Communication}
\label{sec:communication_sensing}

\begin{figure}[t]
	\centering
	\includegraphics[width=\columnwidth]{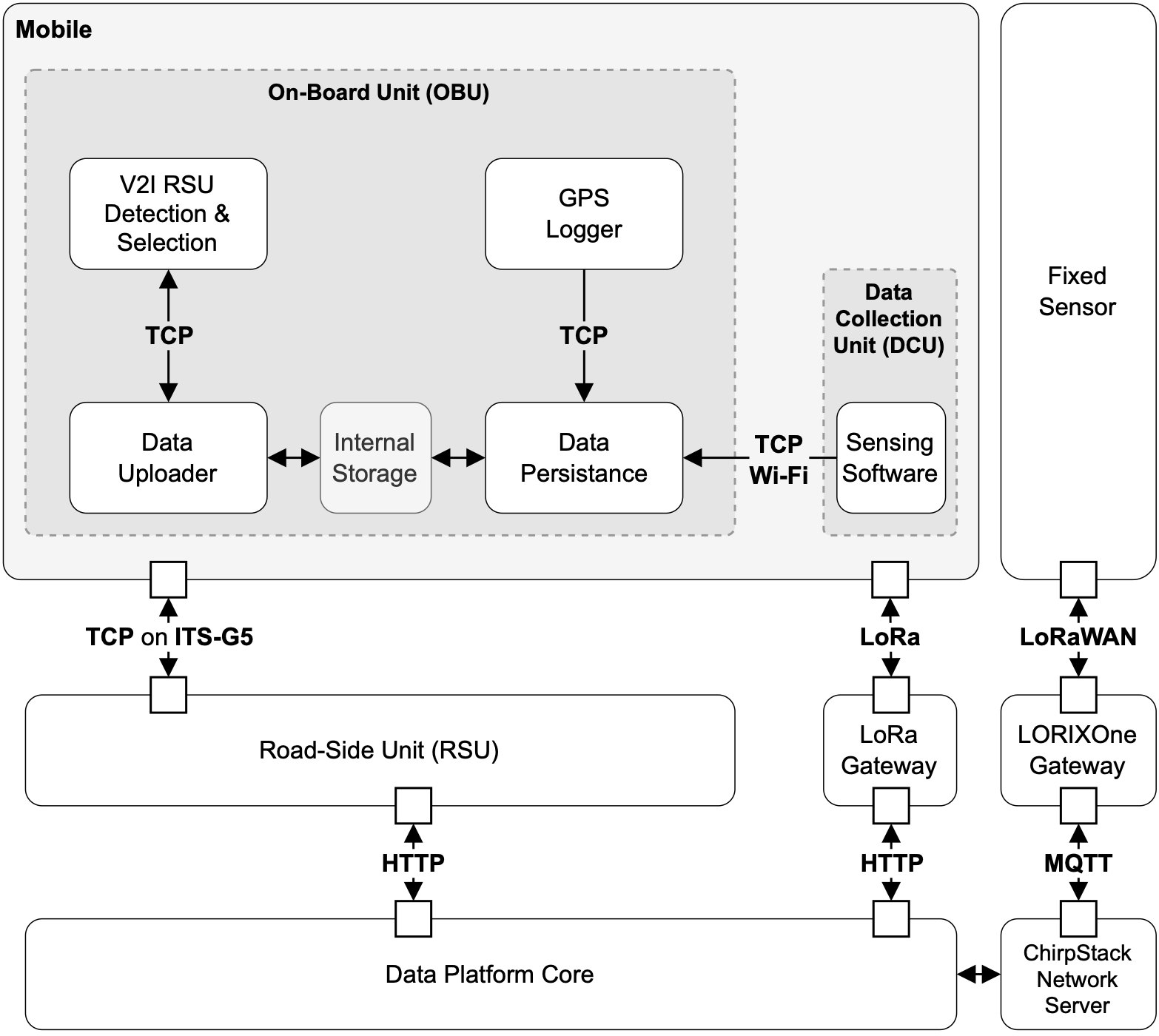} 
	\caption{Sensing to Infrastructure Communication Architecture.}
	\label{fig:SensingToInfra}
\end{figure}

Data generated from the various sensing modules is acquired using different technologies depending on the data source (Figure \ref{fig:SensingToInfra}). The data originating from \ac{dcu} equipment installed on the vehicles can be sent to the \ac{obu} via a wireless WiFi network through an open \acs{tcp} socket. The information is then persisted in a \ac{lifo} queue kept on the filesystem\footnote{www.pypi.org/project/persist-queue}, which aggregates all the data points produced by the different sensors and services active on the bus. Keeping the queue in the filesystem preserves it in the event the bus powers off before communication with the infrastructure is possible.

The \ac{dcu} is constantly connected to the aforementioned WiFi network and transmits new data points every $x$ seconds (this is configured and is set at 3 secs). Additionally, the \ac{obu} itself regularly generates GPS location information and various relevant event logs, which are also queued to be sent to the infrastructure.

\ac{v2i} micro services present in the \ac{v2i} network are continuously polled until a suitable known \ac{rsu} is detected within range, at which point the data queue is compressed, encrypted with \ac{pgp}, and sent via \acs{tcp} through the ITS-G5 or 5G interface. It is then deleted from the \ac{obu} upon acknowledgement of successful integrity verification at the destination (MD5 hashsum). 

Once in the \ac{rsu}, the elements in the queue are processed by a multi-threaded pool of consumers that decode the information and, when applicable, build a standard FIWARE NGSI-LD\footnote{\url{www.fiware-datamodels.readthedocs.io/en/latest/ngsi-ld-howto}} JSON payload that is then sent to the data platform. This software stack supports the collection of several types of data, from multiple sources, and at any desired measurement frequency, thus allowing for future integration with other sensors and different use cases.

Concurrently, the \ac{dcu} modules also make use of their LoRa capabilities to communicate directly with the infrastructure and transmit sensing data to the LoRa gateways installed in the city. These transmissions include the latest measurement information encoded in a custom payload, and occur approximately every 2~minutes and 20~seconds, so as to comply with duty cycle restrictions referring to the use of LoRa frequency bands in the EU. This approach introduces a degree of communication redundancy, and is especially useful to collect up-to-date data in routes with low \ac{rsu} coverage (as opposed to obtaining the stored data later, when a \ac{rsu} is encountered), and in deployments where the \ac{dcu} is the only equipment installed in the vehicles (this is the case of garbage collection vehicles in the city). Once received by a gateway, LoRa frames are decoded and sent to the data platform in the form of NGSI-LD objects, as previously described.


For the sensors using LoRaWAN, the end devices join the network via \ac{otaa}, and communications (both uplink and downlink) are managed by the ChirpStack\footnote{\url{www.chirpstack.io}} Open-Source LoRaWAN Network Server Stack hosted in the ATCLL core. Uplink frames are provided via the HTTP integration to a NodeJS microservice that forwards them to the cloud, which decodes them and returns the measurements for inclusion in the data platform.
Downlink payloads are received by another NodeJS microservice that performs the necessary Network Server API interactions to enqueue them.

\subsection{Vehicle-to-Vehicle and Vehicle-to-Infrastructure}
\label{sec:communication_v2x}

One approach for \ac{v2v} and \ac{v2i} communication is performed using IEEE 802.11p (WAVE), supporting the ITS-G5 communication standard. This implementation uses PC Engines APUs running a standard Linux distribution; therefore, we developed an additional kernel patch to allow these entities to access necessary \acs{ocb} features, making it possible to communicate correctly in the desired frequency range following the ITS-G5 communication standard. Notice that 5G communication is also used, and that, additionally, 2 nodes already communicate through C-V2X.

In order to have information about the surroundings of each node and current network topology, both \acp{obu} and \acp{rsu} need to announce themselves in the correct frequency channel. This announcement is made with \acp{cam} sent in broadcast to every neighboring station. These messages are periodic and contain information related to the position, speed, heading, and status of the sender entity. When an application detects relevant traffic events (e.g. accidents, dangerous end of the queue, emergency vehicle approaching), it can send \acp{denm}. The event can be characterized by an event type, a position, a detection time, and a time duration. The main goal of \ac{denm} transmission is to alert road users about that event.

\begin{figure}[t]
	\centering
	\includegraphics[width=0.8\columnwidth]{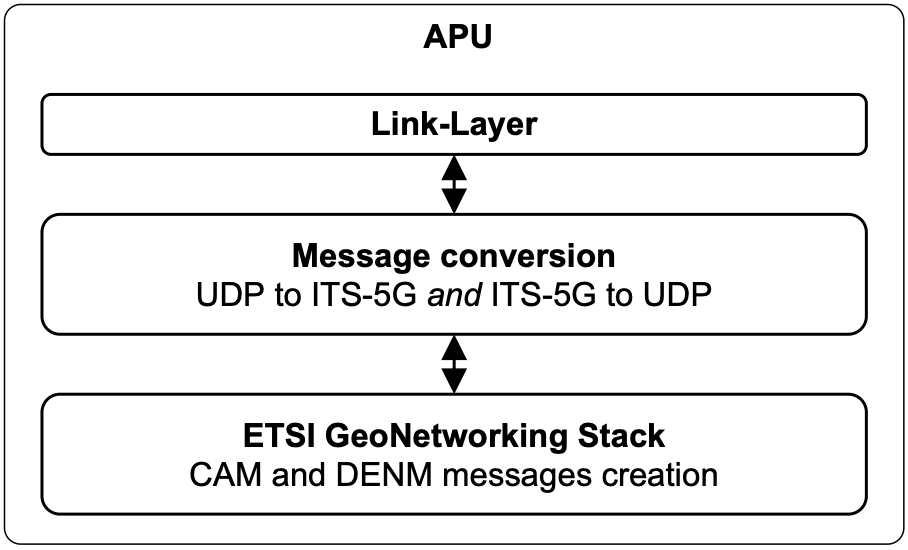} 
	\caption{APU GeoNetworking stack overview.}
	\label{fig:gn}
\end{figure}

The creation of these messages is performed in layer~2 and is done using a GeoNetworking library~\cite{GN} that also includes an ASN.1~UPER encoder. An additional link-layer entity was developed, transforming simple messages (UDP) to standardized messages (ITS-G5), and performing also the opposite message transformation, as shown in Figure~\ref{fig:gn}.


In order to provide Internet to the users using the WiFi \ac{ap} in the \acp{obu}, a connection manager is responsible for choosing in real-time the best ITS-G5 \ac{poa} to assure IP connectivity between the \ac{obu} and the infrastructure. Thus, it needs to determine if there is any ITS-G5 \ac{poa} available to decide if it should be used or a cellular technology instead (4G or 5G). In the case of one or multiple ITS-G5 \acp{poa} in its communication range, it needs to determine which one should be used.

Several network connection management solutions have been proposed in the literature. However, most of them do not consider high-mobility scenarios. In~\cite{bib:Dias12} it was proposed a \ac{vcm}, which can operate in high-mobility, making the best decision possible for the connection selection based on different metrics such as speed, location, \ac{rssi}, and the expected contact time. Based on this work, the authors in~\cite{pbcm} presented an upgraded approach. Besides taking into account the same base metrics as the previous work, which only chooses the best network available at a specific instant, this new approach proposes a predictive connection manager that can decide, in advance, the best network and technology for the attachment. 

A straightforward implementation of the connection manager is to establish a connection to the \ac{poa} that presents the best \ac{rssi}, and settle some thresholds to determine when it should consider changing to another \ac{poa} with a better \ac{rssi} level or not. A more efficient version from the previous mechanism is to add the GPS position and heading from the vehicle. This avoids connecting to an \ac{rsu} that, besides having the best \ac{rssi} currently, is located on the other side of the road, so the \ac{rssi} would decrease rapidly while driving in the opposite direction. Such solution is the one that is currently implemented in the \ac{atcll} network, with the addition that it is capable of changing between an ITS-G5 and 5G connection when there are no \acp{rsu} in the surroundings of the vehicle. Current work addresses the improvement of this mechanism, considering other metrics and \ac{sdn} capabilities in the network.

\subsection{Software Defined Vehicular Network}

The \ac{sdn} solution implementing the vehicular network is based on an architecture with proactive mobility detection, where the handover is detected before it happens, even if there is no traffic. For this purpose, the \acp{rsu} themselves are part of the \ac{sdn} topology by also being \ac{sdn} switches. This way \acp{rsu} can communicate to the controller signalling information concerning the nearby moving elements. 

\begin{figure}[t]
\centering
\includegraphics[width=1\linewidth]{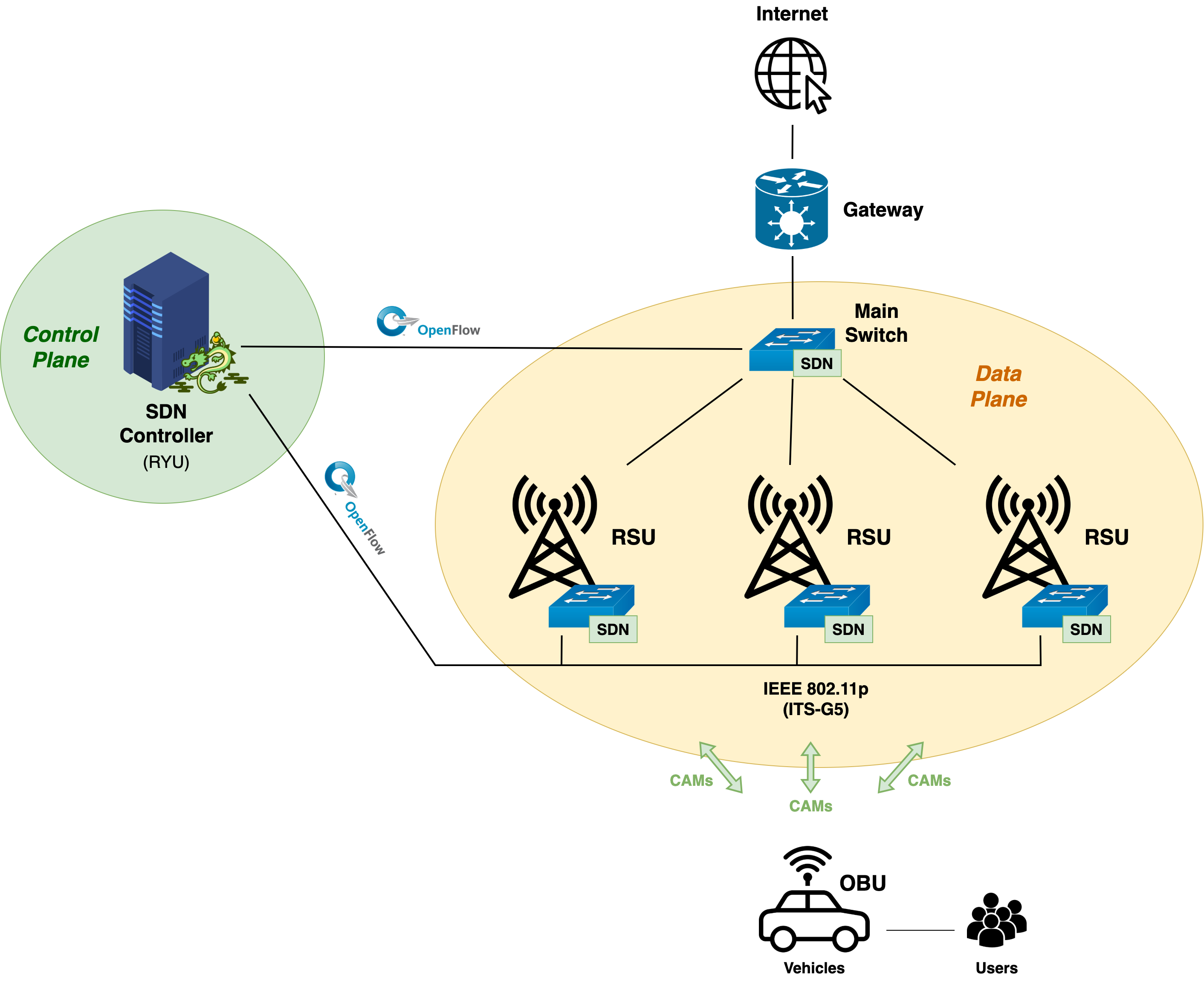}
\caption{Software Defined Vehicular Network architecture. \cite{DDias}}
\label{fig:proactive_arch}
\end{figure}

To offer proactiveness in the handover process, the most suitable class of messages are \acp{cam} because they carry signalling information such as location, heading, speed and type of vehicle. This information is very relevant in the controller because it allows it to have an idea of the movement of the \acp{obu}, thus allowing the prediction of handovers even before they happen. Figure \ref{fig:proactive_arch} depicts the architecture \cite{DDias}.

A custom protocol (OBUInfo) was introduced whose ethertype number is 0xBBBB (which is not assigned by the IEEE 802 standard). The protocol was developed to be able to differentiate between custom OBUInfo messages (containing the contents of \ac{cam} messages plus the \ac{rssi} forwarded by the \acp{rsu}), which represents control traffic, from IP packets that represent data traffic. Therefore, the controller can separate both, so there is no need to send control messages within data traffic. The algorithm running in the controller which selects the preferable \ac{rsu} to communicate with the \ac{obu} is based on the same conditions used in the connection manager described in the previous subsection. 

\subsection{People-to-Infrastructure}
\label{sec:communication_people}


\Ac{p2p} and \ac{p2i} communication is performed using an ITS station supporting communication technologies, such as WiFi or cellular technologies, including LTE and 5G. This implementation uses smartphones. The choice of the smartphone is motivated by its ubiquity, high programmability, and reduced cost compared to other solutions. We developed an additional application to allow the \acp{vru} to communicate with the infrastructure, making it possible for \acp{vru} to communicate information about them and receive notifications from the infrastructure about potentially dangerous situations (\emph{e.g.}, potential crashes).

The information obtained from the ITS station allows the announcement of the necessary awareness information about \ac{vru} — location, altitude, heading and speed, but also orientation, direction, size and weight class and \ac{vru} profile. To obtain this information set, the information from the smartphone from inertial sensors (accelerometer, gyroscope, magnetometer, and orientation sensors), from the cellular signal strength and WiFi, Bluetooth information and GNSS/GPS sensor is fused using the Fused Sensor Provider API from Google\footnote{\url{https://developers.google.com/location-context/fused-location-provider}}. The announcement of this awareness information is made with \acp{vam} sent by the smartphones. Similar to \acp{cam}, the creation of these messages is done using an ASN.1~UPER encoder, and they are transmitted to the infrastructure through access technologies, namely WiFi, LTE and 5G.

The developed application integrated with the sensors makes the smartphone an ITS station capable of similar basic functionalities of an \ac{obu}. By obtaining and processing the necessary information to generate periodic messages to make the vehicular network aware of the vehicle and its essential information — such as location — the smartphone can act as a low-cost \ac{obu}. The developed application predicts this situation by changing the station type of the produced message to a vehicle if it detects that the speed of the smartphone is not compatible with a \ac{vru} walking. 

Finally, the infrastructure is capable of delivering Internet access to people throughout the city. The \acp{rsu} also have a WiFi link that behaves like a typical \ac{ap} providing internet access to the users directly connected to them. 









\section{Computing/Edge Approach}
\label{sec:edge}

This section presents the edge approach for data computation and the multi-sensor data fusion and detection.

\subsection{Edge support}
\label{sec:edge_support}

Every edge point has available, apart from the communication equipment, an Nvidia Jetson Nano or Jetson Xavier with a powerful GPU for graphic-intensive processing, and a Raspberry Pi 4 for the deployment of lightweight services. This edge computing capacity is leveraged to maintain several data collection and processing services running on the different \acp{sbc} installed at each edge node. 
Some examples include:

\begin{itemize}
    \item Decoding and parsing of the traffic radar message protocol;
    \item Real-time CUDA-assisted object detection in a video stream (detection of different vehicles, people and objects visualized through video cameras);
    \item Decoding and parsing of ETSI ITS information in \ac{cam} and \ac{denm} messages, and interactions with kernel subsystems to ascertain the respective \ac{rssi};
    \item Periodic calculation of vehicle counts and velocity averages for different types of vehicles within specified time frames (separate process for traffic radar, video cameras and \ac{cam} data);
    \item Data fusion processing using traffic radar, video cameras CAM data.
\end{itemize}

Current resource utilization metrics and load averages show that there is sufficient headroom for plentiful future additions. Given that some services must consume the data produced by others in order to work, a comprehensive solution for publishing and accessing data is crucial to ensure the effectiveness of the edge computing architecture, while minimizing resource consumption overhead and latency for timing-sensitive workloads.

\begin{figure}[t]
\includegraphics[width=\columnwidth]{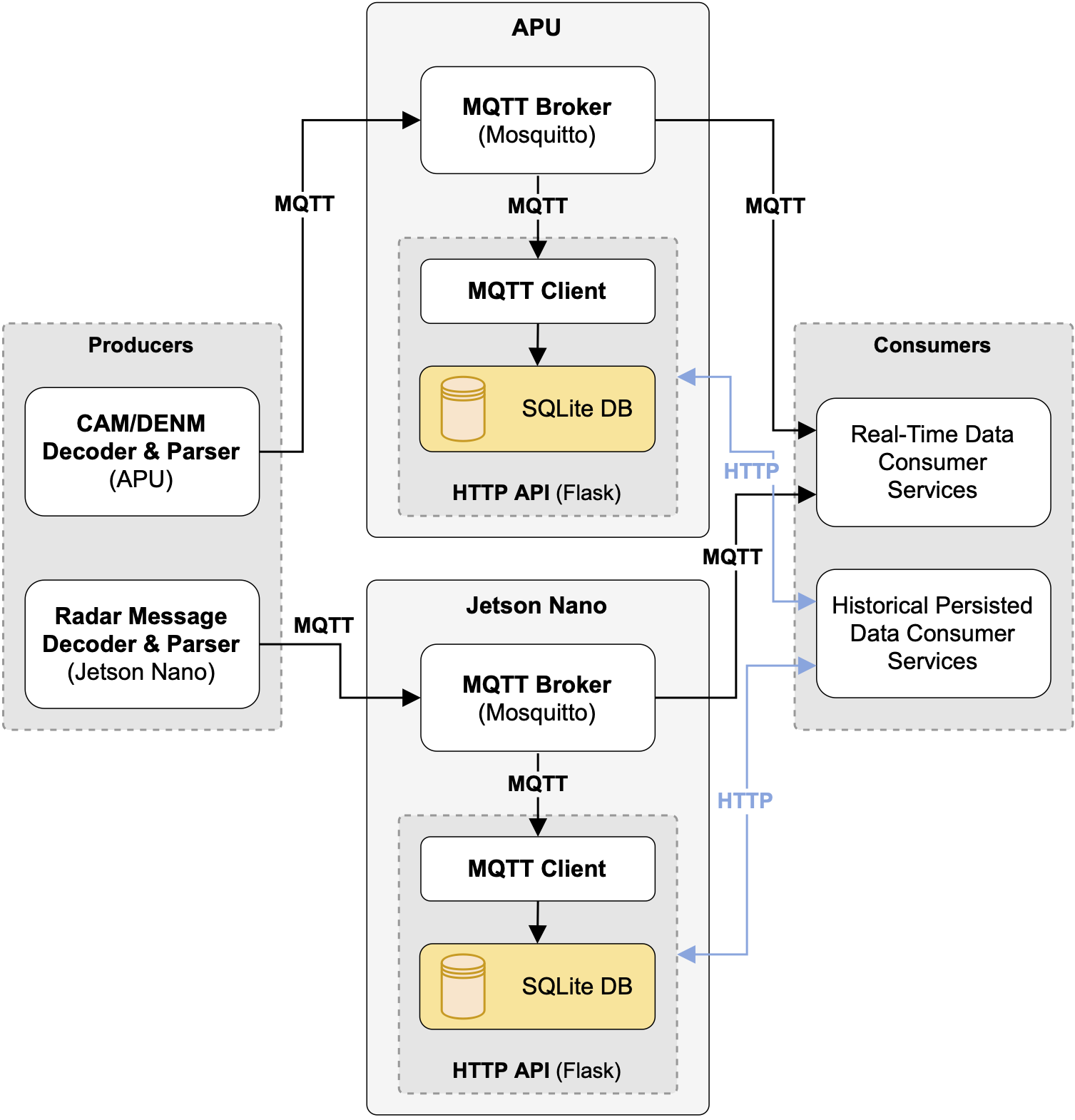} 
\caption{Edge computing data access strategy.}
\label{fig:edge_data_access}
\end{figure}

Two separate types of data access are supported (the edge strategy for data access is depicted in Figure~\ref{fig:edge_data_access}):

\begin{itemize}
  \item \textbf{Real-time information stream} - Each edge node hosts a Mosquitto \ac{mqtt} broker to be used by all the equipment and services installed on that particular node. As such, a publisher/subscriber model is used, where producers publish new data points to specified topics, and consumers subscribe to the relevant topics and process each new message;
  \item \textbf{Persisted historical information} - Each edge node also hosts an SQLite database configured to persist the real-time data published to a specified list of \ac{mqtt} topics for a short time period (with a default value of 24 hours).
  This functionality is aimed for data processing services which require complex queries and a complete history of relevant data-points, such as data fusion algorithms.
  Access to the persisted information is achieved through the use of a REST API developed for this purpose, since SQLite does not natively support multiple concurrent connections.
  The choice of SQLite as \ac{dbms} over other more feature-rich alternatives is focused on limiting excessive resource overhead. 
  Several \acs{dbms}-specific optimizations were configured in order to greatly reduce read and write times.
\end{itemize}

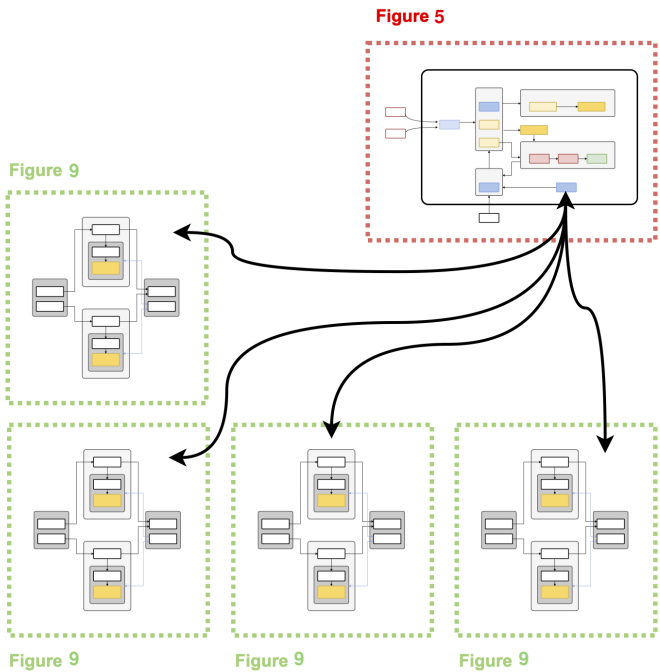
\begin{figure}[t]
\begin{overpic}[width=\columnwidth]{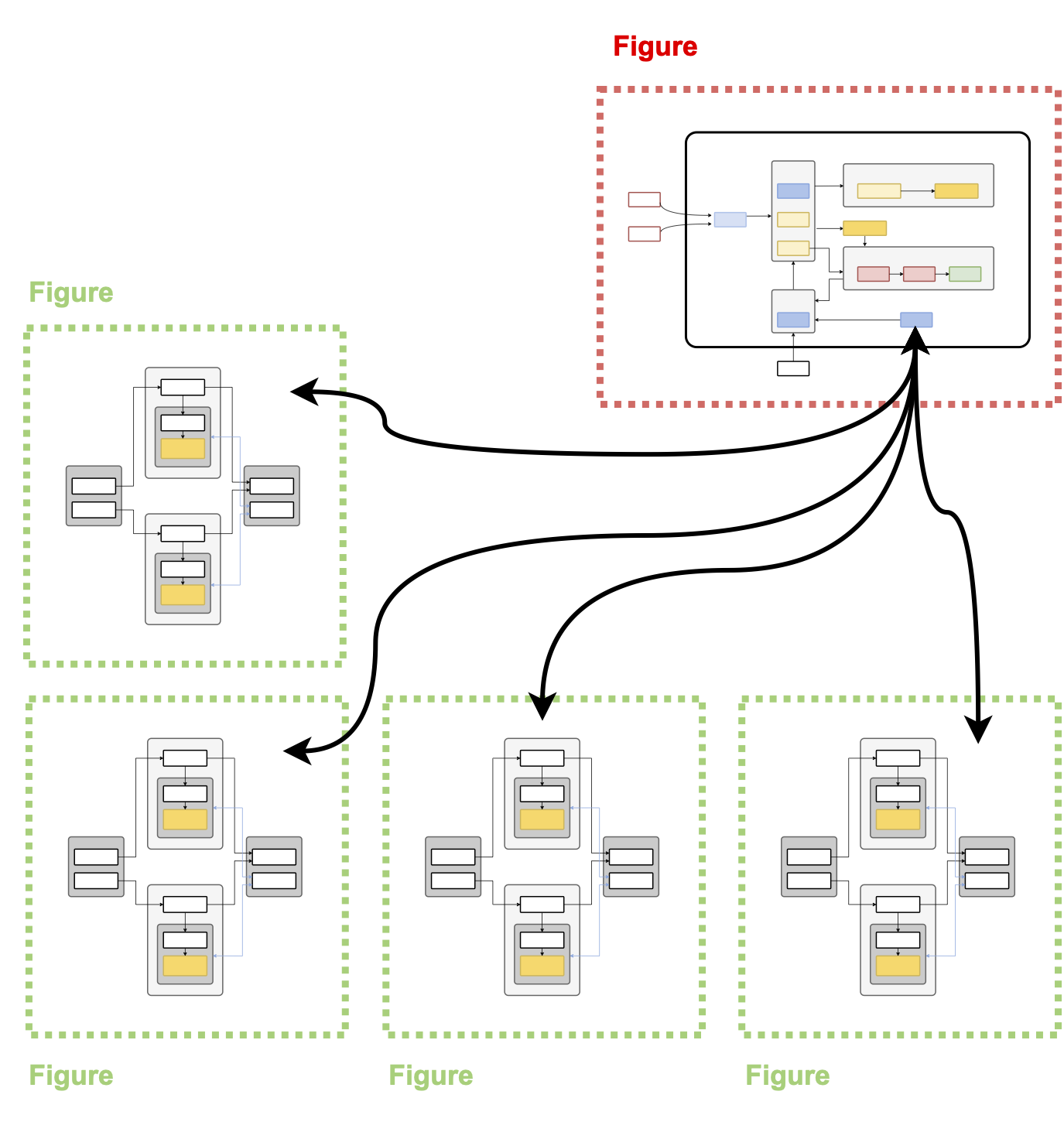}
    \put (11, 3.5) {\scriptsize\sffamily\textcolor{figgreen}{\textbf{\ref{fig:edge_data_access}}}}
    \put (43, 3.5) {\scriptsize\sffamily\textcolor{figgreen}{\textbf{\ref{fig:edge_data_access}}}}
    \put (75, 3.5) {\scriptsize\sffamily\textcolor{figgreen}{\textbf{\ref{fig:edge_data_access}}}}
    \put (11, 73) {\scriptsize\sffamily\textcolor{figgreen}{\textbf{\ref{fig:edge_data_access}}}}
    \put (63, 95) {\scriptsize\sffamily\textcolor{figred}{\textbf{\ref{fig:data_platform}}}}
\end{overpic}
\caption{Edge computing data access overview through the cloud.}
\label{fig:edge_mqtt_overview}
\end{figure}

Additionally, a separate centralized \ac{mqtt} broker instance running in the cloud serves as an \ac{mqtt} bridge between the individual edge node brokers and any cloud services, dashboards, and frontend UIs that also require access to \mbox{real-time} edge data (albeit with slightly increased latency when compared with edge consumers closer to their respective data sources). This is depicted in Figure~\ref{fig:edge_mqtt_overview}.

\subsection{Automated Deployment}

Deploying applications to an infrastructure of this size and complexity presents a number of challenges relating to automation and dependency management, among others. These types of deployments can become increasingly harder to manage and maintain, given the high number of nodes and the fact that they are heterogeneous in terms of system architecture, operating system version, and other factors. Moreover, dependency version incompatibilities may occur between services that run on the same compute node.

To address these issues, applications are deployed as Docker containers, which package the application code with all of its dependencies and isolate them from the rest of the system. 
Using this approach, applications can be easily deployed on any Docker-enabled machine in either the cloud or the edge, while maintaining low levels of virtualisation overhead. This also greatly simplifies the deployment process for any third-party services that use the infrastructure.

Alternatively, in cases where a transition to a containerised environment has not yet been completed or has been found to be unfeasible, applications can also be deployed as more traditional $Systemd$ services. This type of deployment uses Ansible playbooks to ensure a reliable, automated process which includes the installation of the required dependencies, and can be run concurrently on multiple target nodes.

Additionally, the project also leverages CI/CD methodologies and the Gitlab platform in order to automatically deploy updated versions of an application to the respective compute nodes each time a new commit is published to specified production branches. This process includes the automated building of Docker image variants for each CPU architecture (x64, armv8, armhf) using Docker BuildX and QEMU.

Current work addresses the possible integration of this infrastructure with a more advanced service orchestration framework such as Kubernetes.

\subsection{Multi-sensor fusion and detection}
\label{sec:edge_fusion}

Vehicles, bicycles, boats, people and other objects can be detected by multiple sources. The usage of several data sources improves the reliability by eliminating the dependency on just one type of sensor.

Information about vehicles can come from vehicles OBUs, while information about people can come from their smartphones. In addition, both vehicles and people, and also bicycles, boats and other objects can be detected through video-cameras, traffic radars, LiDAR and the analysis of the electromagnetic spectrum.

Video-cameras are scattered throughout the city. Using a video camera and a Nvidia Jetson, a detection process of objects can be implemented. It starts by creating a \ac{rtsp} stream, established to transmit the most recent frame, with the camera integrated in the Smart Lamp Posts. 
During this process, in order to detect the pedestrians effectively, each time the Jetson device is not already busy processing a frame, it fetches the most recent frame from an internal queue, and then applies a set of object detection functions based on a deep learning approach — by using a pre-trained YoloV3 Model\footnote{https://pjreddie.com/darknet/yolo/} or the Nvidia DeepStream SDK Object Detector\footnote{https://developer.nvidia.com/deepstream-sdk}. Both models are optimized to the Nvidia Jetson hardware — the DeepStream SDK being created specifically for the Jetson device, and the YOLOv3 with the usage of the module TensorRT\footnote{https://developer.nvidia.com/tensorrt}, also from Nvidia. Then, at the end of each cycle, the result is published in the edge \ac{mqtt} broker, indicating the presence of at least one object, such as pedestrians, vehicles, bicycles, etc.

As mentioned, Smart Lamp Posts contain $smartmicro$ traffic radars\footnote{https://www.smartmicro.com/} that detect and track objects. This type of radar gives information about position, direction, speed and type of vehicle (classification in light vehicles, heavy vehicles and motorcycles/bicycles) of each detected vehicle. All this information is sent to the Nvidia Jetson that receives, decodes and publishes the data on a \ac{mqtt} broker, following the edge data strategy described in the previous subsection. 
With this information it is possible, in real time, to display the traffic situation, send messages to neighbouring \acp{rsu} with traffic information, and make statistic reports: at what time there are peaks of traffic, what are the types of vehicles that circulate in this area, and what is the average velocity and acceleration of those vehicles. 

The level of accuracy of the traffic radar to detect the type of vehicle is around 80\%, and its intrinsic restriction of not being able to differentiate smaller objects (people, bicycles, and motorcycles), limit the applicability of this sensor for some use cases. In order to cover such type of detection, the \ac{atcll} infrastructure also has LiDAR sensors attached to the Smart Lamp Posts. LiDAR has a much higher resolution, allowing to obtain better results in terms of location and classification of vehicles. However, the amount of data produced by the sensor, due to the creation of point clouds,  means that the output data rate is much higher in comparison to the traffic radar (around 6500 times higher). The usage of the LiDAR for this application is one more example of the importance of the edge processing. By processing the point clouds closer to the sensor avoids a large occupation of the network channel, in comparison to if the data would be processed exclusively in the core. Instead, the core consumes the output metadata of the edge processing, also following the edge data strategy described in the previous subsection. 

Another alternative for the detection of objects, more specifically the detection of flows of people, is through smartphones and their connectivity - using a technique known as WiFi probing. The WiFi \acp{ap} present in the Smart Lamp Posts and buildings can be turned into monitoring mode to be able to detect the different WiFi terminals around and infer the number of devices connected to WiFi. With this inference, it is possible to detect people carrying their own smartphones or other equivalent devices.

\section{Data persistence, processing and sharing}
\label{sec:data_mgmt}

\subsection{Data persistence and processing}
\label{sec:data_mgmt_mechanisms}
The main services used to manage and persist the data are the \emph{Processing} server and the ClickHouse database. The \emph{Processing} server is responsible for a preliminary processing before the data reaches Orion, more specifically finding the road segment of each received point, and then persisting it on the respective table, depending on the FIWARE-Service attribute and on the data type. For the persistence of the data collected through the sensors and communication devices in the infrastructure, \ac{atcll} implements a \ac{dbms}, ClickHouse. This database has native compression capabilities that can greatly decrease the databases’ size, while speeding up the analytical queries. The distributed configuration of ClickHouse is also considered straightforward. It is composed of clusters that have a configurable number of shards, independent groups of replicas that store and operate on a part of the stored data. In the current form of the data platform, a single ClickHouse shard is used. However, a cluster configuration is still deployed in order to achieve a greater availability and redundancy that can be obtained with replicas. Expanding the infrastructure and thus increasing the amount of data, the cluster size can be increased, adding a new shard with the same number of replicas. 

\subsection{Data sharing}
\label{sec:data_mgmt_sharing}

The Orion Context Broker is accessible from the outside, which is highly configurable and of high performance. It allows both the reception of data into the platform, and also the distribution to the consumers in real time. Using a token-based security layer on top of the broker, managed by the \emph{Authentication} server, the information is kept on the platform as consistent as possible, while only giving access to authorized entities.

On top of the real-time information being transported through the \ac{atcll} platform, the \emph{Processing} server has also available several public APIs intended to make the data available for the dashboard, as well as a historical API, secured by the token authentication method, that can be used to obtain persisted data from any service and type of any time interval.

\begin{figure}[t]
\includegraphics[width=\columnwidth]{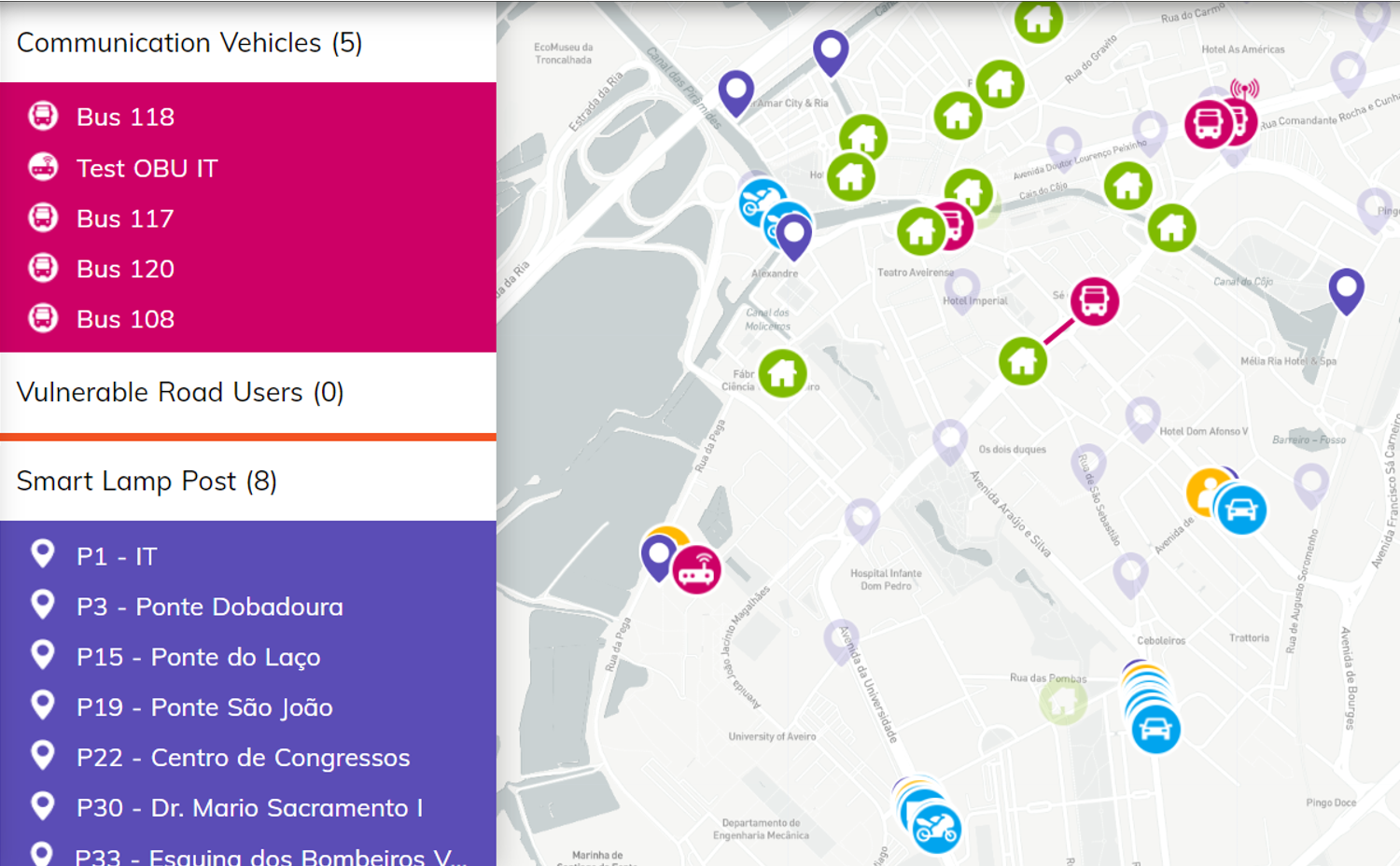} 
\caption{Screenshot of the ATCLL dashboard showcasing real-time data.}
\label{fig:realtime}
\end{figure}

Besides the public Orion broker, to keep the connection with the outside, we use two additional brokers on the platform. Firstly, we use Kafka\footnote{https://kafka.apache.org/} to distribute all the real-time data that goes through the public broker inside the network. All the services inside the network that need to have the data in real time should use this Kafka broker in order to lighten the Orion load. The other one is a \ac{mqtt} broker that serves as a bridge to all the edge devices, making available all the edge computing data in one internal place only, allowing central computing and cloud data persistence (Figure~\ref{fig:edge_mqtt_overview}).


\section{Results}
\label{sec:results}

This section presents two main types of results: (i)~examples of data collected from the sensing and communication devices of the \ac{atcll} infrastructure, (ii)~and performance results.

\subsection{Examples of data collected}
\label{sec:results_examples}

The \ac{atcll} dashboard is a key piece of the platform which enables the visualization of the collected data (in real-time and from historical records), geographically, and in plots for representation over time.

\begin{figure}[t]
\includegraphics[width=\columnwidth]{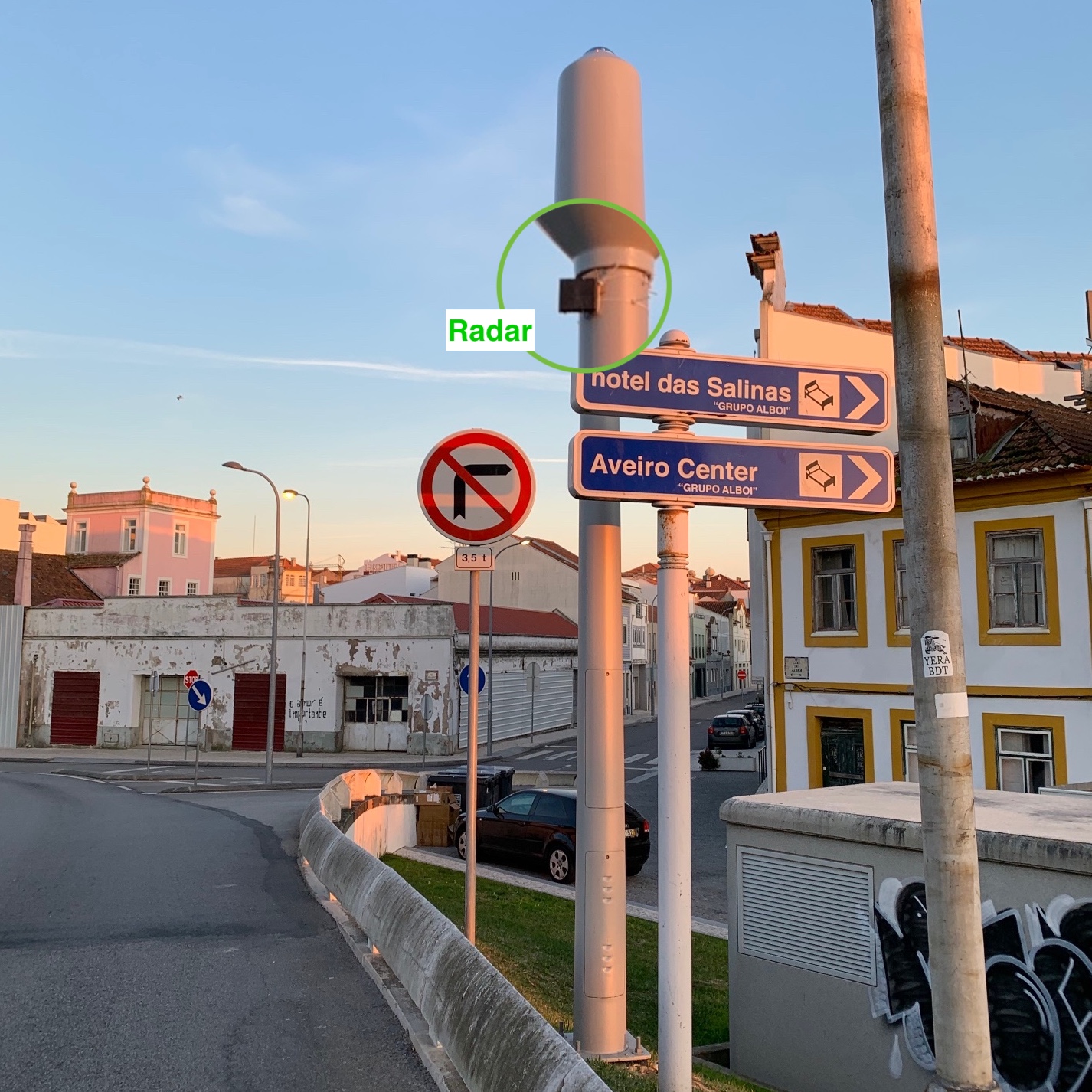} 
\caption{Traffic radar localization — Dobadoura bridge, Aveiro}
\label{fig:ponte_dobadoura}
\end{figure}

An example of the visualization of real-time data is depicted in Figure~\ref{fig:realtime}: the nodes in blue are vehicles detected by the traffic radar, which is able to distinguish three classes of objects as of light vehicles, heavy vehicles, and others including pedestrians and \mbox{two-wheelers}. The nodes in pink are detected by connectivity (\ac{v2x} or cellular), where pink lines represent the active \ac{v2i} connection between the vehicle and the respective \acp{rsu}. In such an example, as depicted, it is possible to observe that various \acp{rsu} were able to receive \acp{cam} transmitted by the same vehicle. Moreover, the yellow icons represent the detection of objects performed by a video camera, whose target can be of several types: pedestrians (as seen depicted in the figure), bicycles, motorbikes, cars, trucks, among others.

In the dashboard, the user is also able to choose the time ranges to visualize historical data. For instance, to visualize the data related to the number of vehicles that crossed a certain area in a period of time, the user may choose the traffic radar data.

\subsubsection{Traffic radar}

\begin{figure}[t]
\centerline{\includegraphics[width=9cm]{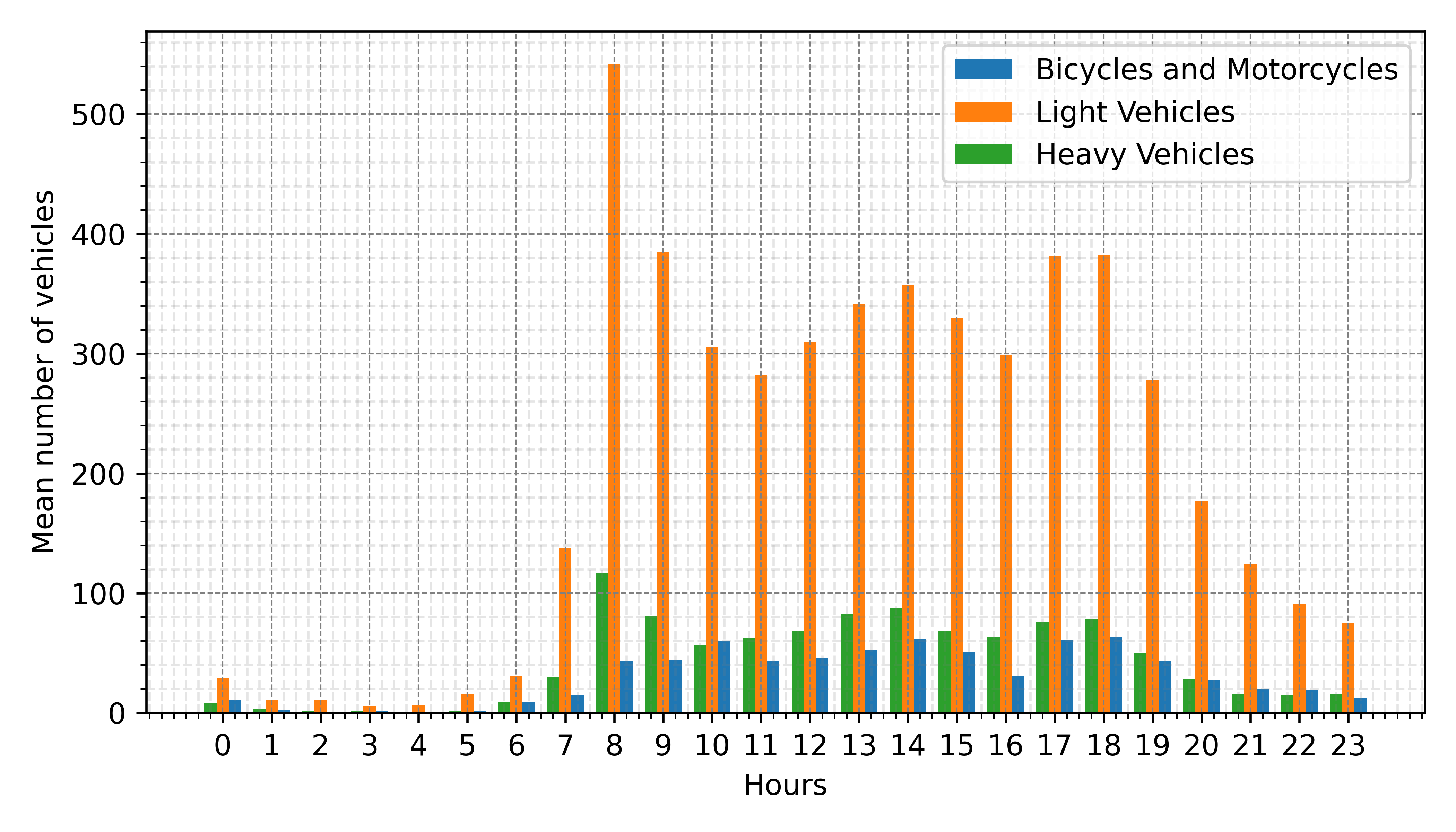}}
\caption{Hourly average of the number of vehicles arriving to the city through Dobadoura (2022/02/07—2022/02/11). }
\label{fig:n_mean_vehicle_P3}
\end{figure}

\begin{figure}[t]
\centerline{\includegraphics[width=9cm]{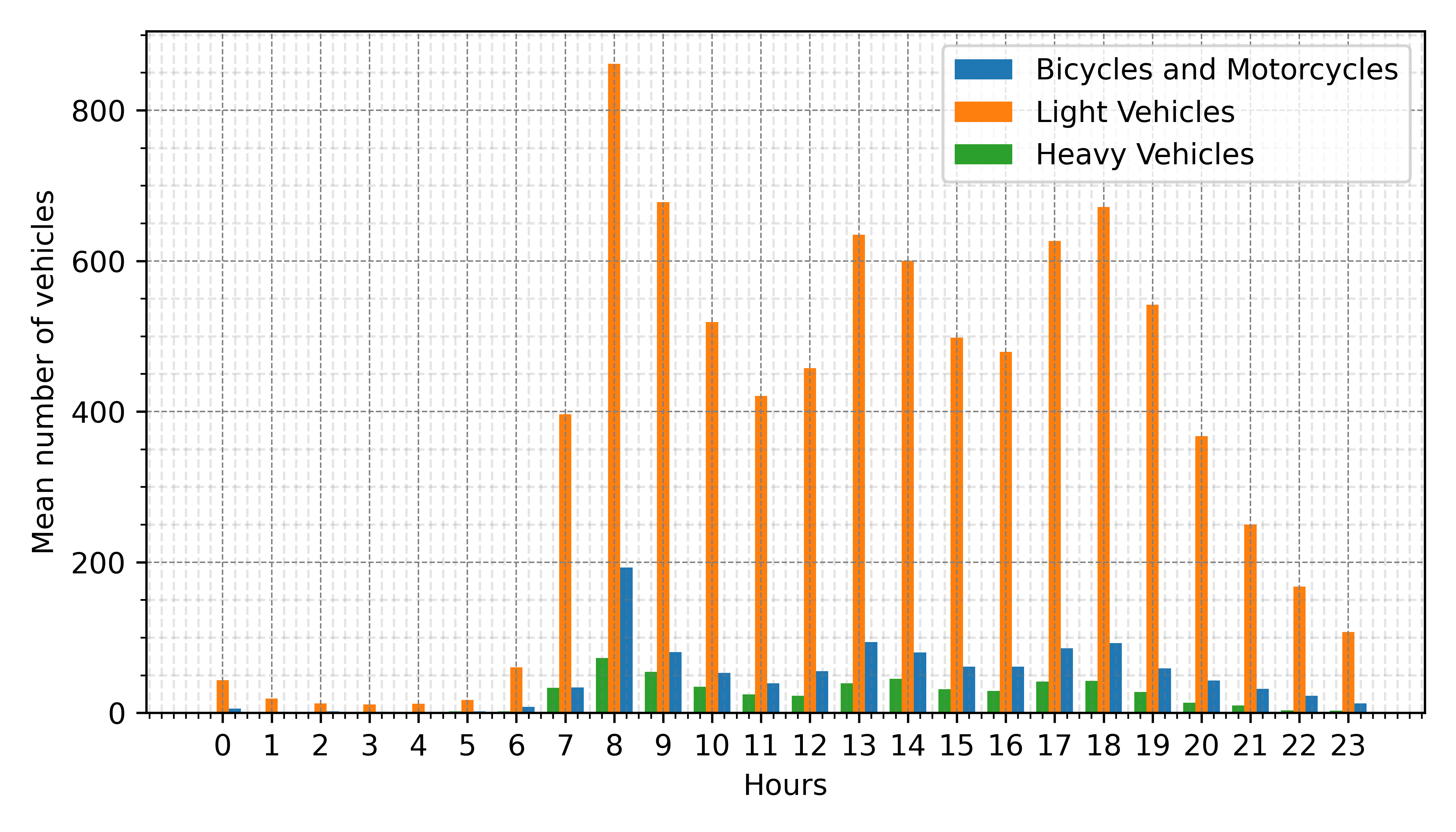}}
\caption{Hourly average of the number of vehicles arriving to the city through \mbox{ISCA-UA} (2022/02/07—2022/02/11). }
\label{fig:n_mean_vehicle_P35}
\end{figure}

\begin{figure}[t]
\centerline{\includegraphics[width=9cm]{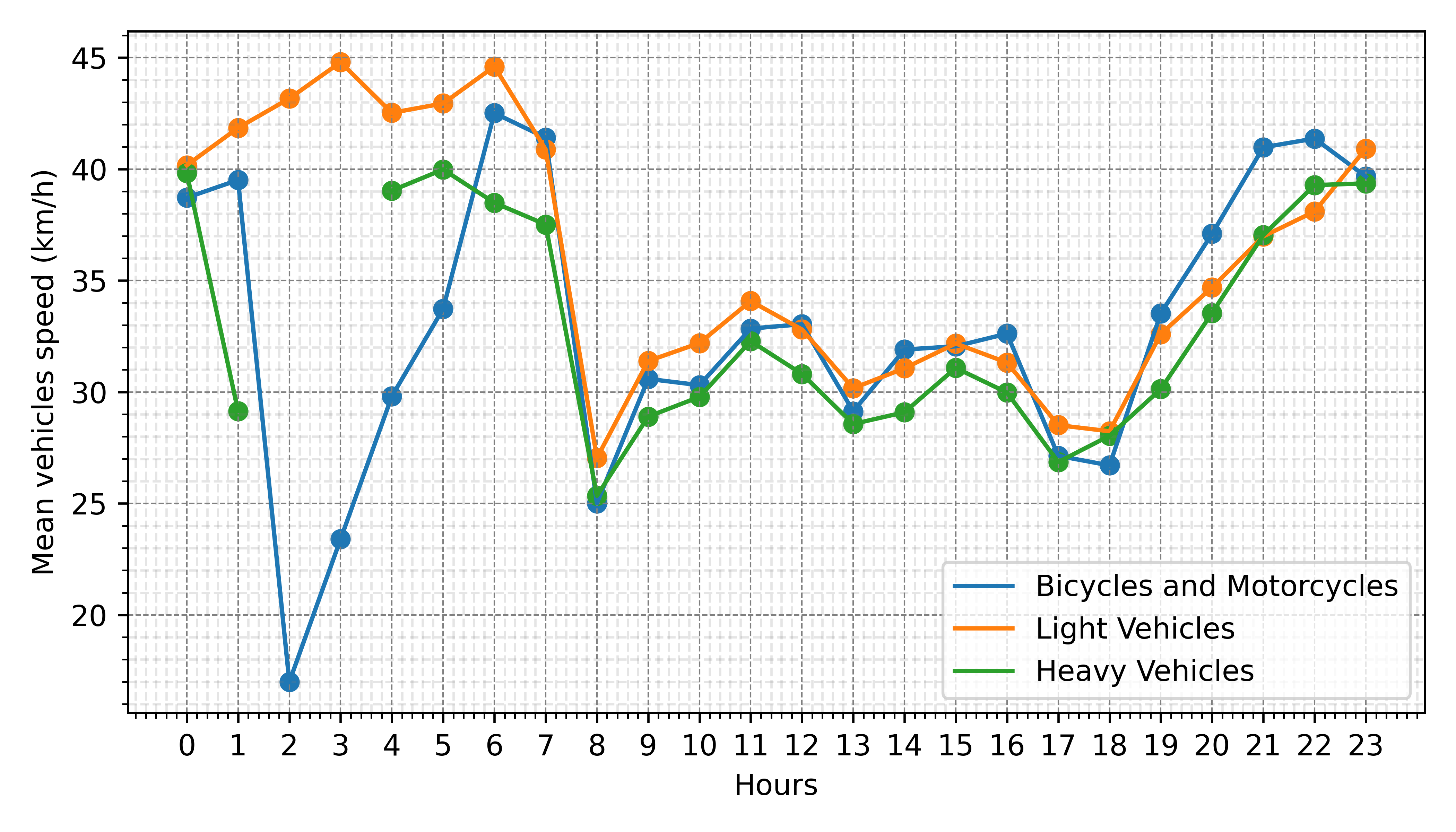}}
\caption{Hourly average of the vehicle's speed entering the city through \mbox{ISCA-UA} (2022/02/07—2022/02/11). }
\label{fig:speed_mean_vehcile_P35}
\end{figure}

\begin{figure}[t]
\centerline{\includegraphics[width=9cm]{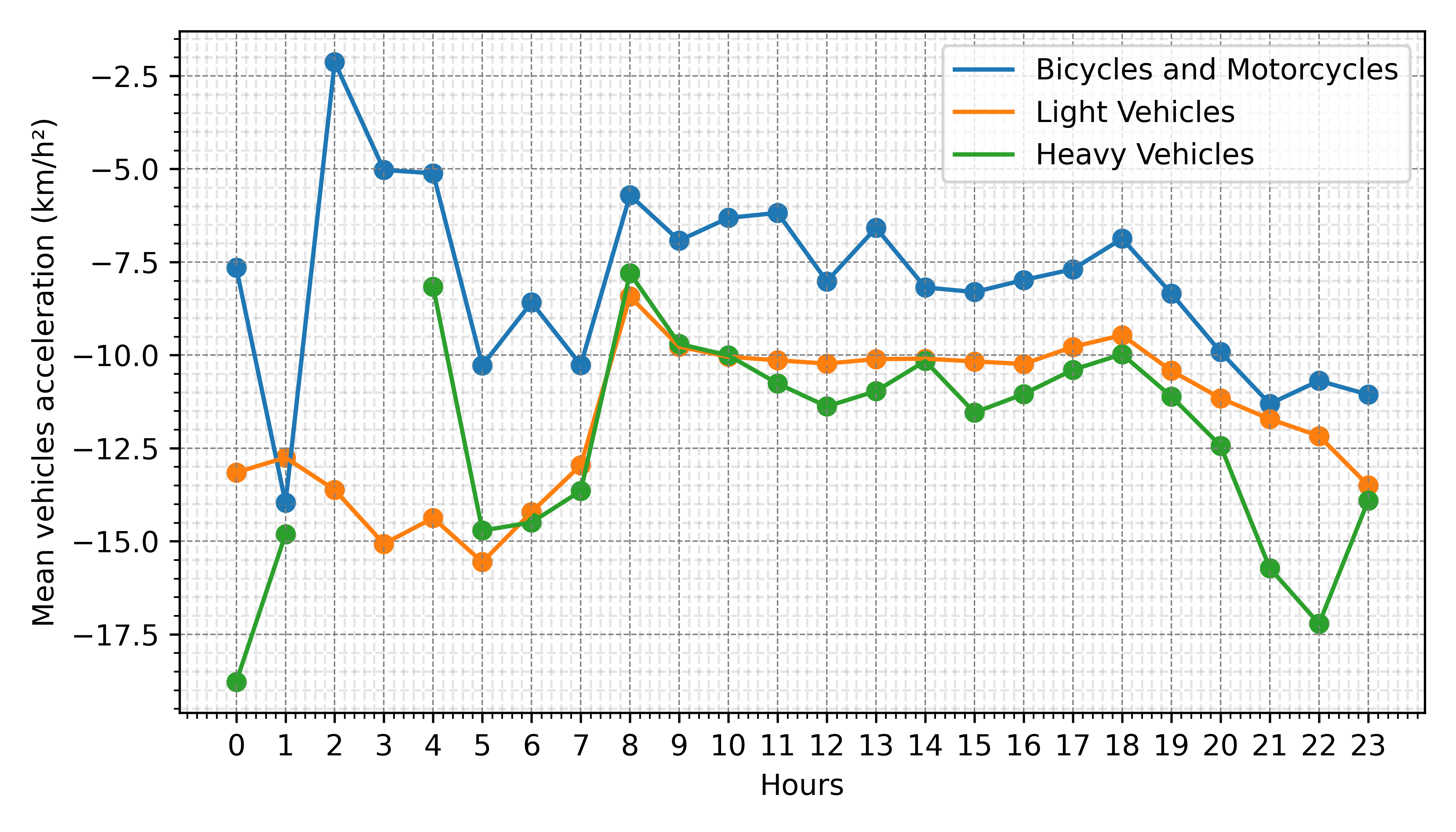}}
\caption{Hourly average of the vehicle's acceleration entering the city through \mbox{ISCA-UA} (2022/02/07—2022/02/11).}
\label{fig:acceleration_mean_vehcile_P35}
\end{figure}

Traffic radars provide the information of the number and types of vehicles per time that cross a certain area, as well as their velocity and acceleration. Examples of radars are the ones installed in Dobadoura bridge (Figure~\ref{fig:ponte_dobadoura}) and in \mbox{ISCA-UA} (Figure~\ref{fig:wb-slp}).
Figures~\ref{fig:n_mean_vehicle_P3} and \ref{fig:n_mean_vehicle_P35} show the average number of vehicles that, in each hour, have crossed Dobadoura bridge and \mbox{ISCA-UA}, on the week between February~7th and February~11th, 2022.
Figures~\ref{fig:speed_mean_vehcile_P35} and \ref{fig:acceleration_mean_vehcile_P35} show the average speed and acceleration of vehicles which entered the city at \mbox{ISCA-UA} on the same time period.

From this data, we can infer several aspects related to the city mobility. Comparing the number of light vehicles entering the city in Dobadoura bridge and \mbox{ISCA-UA}, we observe almost the double number of vehicles entering using the \mbox{ISCA-UA} access. Moreover, we see that the traffic peak time occurs at the same time for both locations, around 8am. Finally, it is possible to correlate the number of vehicles and their speed/acceleration: during daylight periods, as traffic congestion increases, one can observe a reduction in speed values, and a near-constant acceleration in arriving vehicles; on the other hand, in the night periods, no traffic congestion is noticed, leading to a higher variance of speed values, consequently causing oscillating acceleration values.

\subsubsection{Camera}

Another type of data available in the dashboard for the user is the video camera detection. Through the vision framework running in the edge, already described in Subsection \ref{sec:edge_fusion}, the platform is able to accumulate data regarding the detection of objects from different classes. The results presented in this article focus on the flow of pedestrians in the areas where cameras are located.

\begin{figure}[t]
\centerline{\includegraphics[width=9cm]{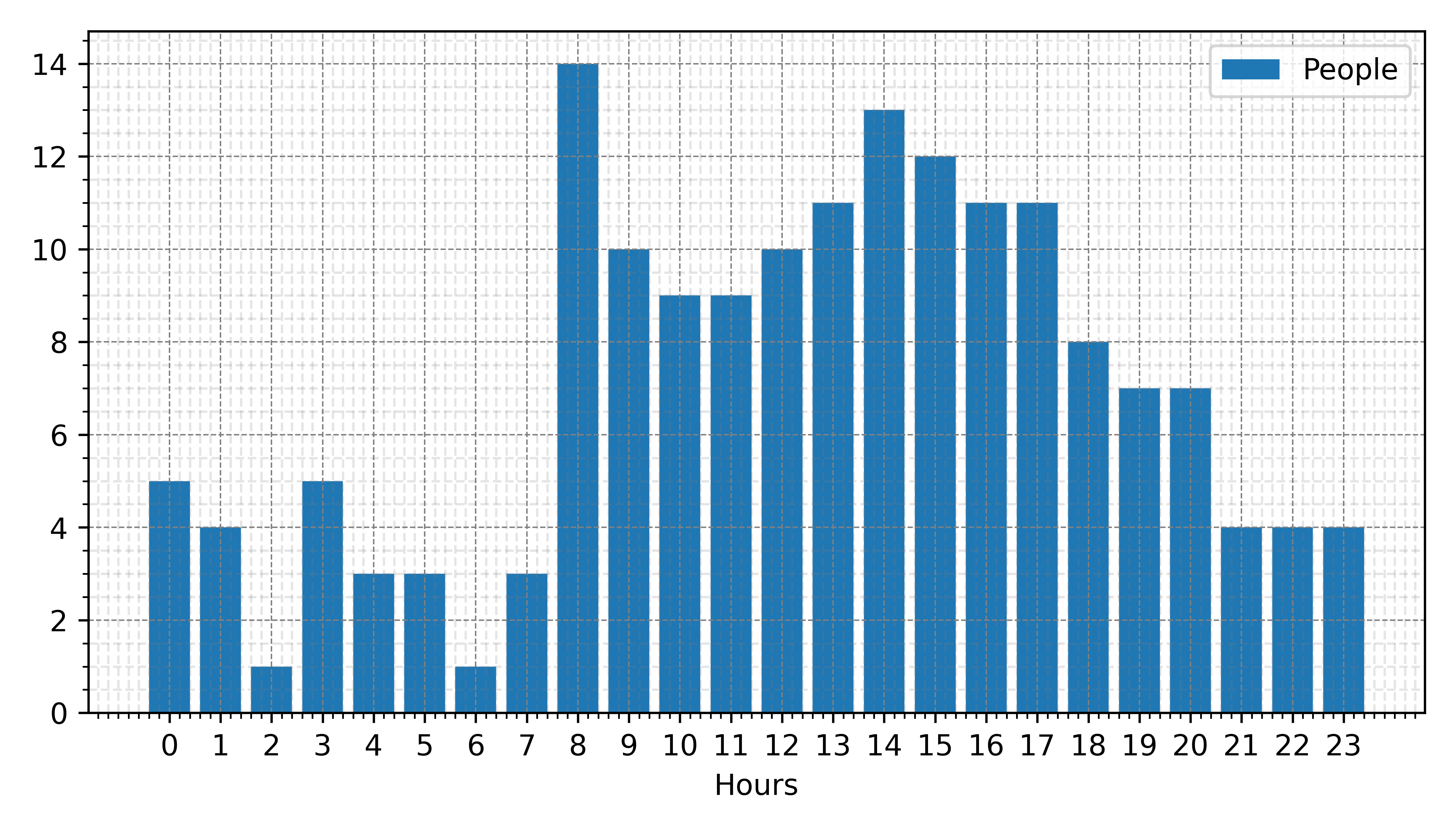}}
\caption{Maximum number of pedestrians observed in a given video frame in the crosswalk located next to the \mbox{ISCA-UA} (2022/02/08).}
\label{fig:n_people_isca_p22}
\end{figure}

In the Smart Lamp Post of \mbox{ISCA-UA}, there is a video camera (Figure~\ref{fig:wb-slp}) capturing a crosswalk located at the end of the road being sensed by the traffic radar. Figure~\ref{fig:n_people_isca_p22} depicts statistics on the detection of pedestrians using that crosswalk. At each instant of the detection, the number of distinct people is counted and the maximum of these values, per hour, is plotted. With this data it is possible to analyze some correlation between Figures~\ref{fig:n_mean_vehicle_P35} and \ref{fig:n_people_isca_p22}. When more people are detected, for instance at 8am, the vehicles' average speed is lower. There are some studies already in place by the municipality of Aveiro regarding this intersection, to evaluate the impact of the crosswalk occupation in the vehicles traffic exploring the information collected by the radar and the video camera.

\begin{figure}[t]
\centerline{\includegraphics[width=9cm]{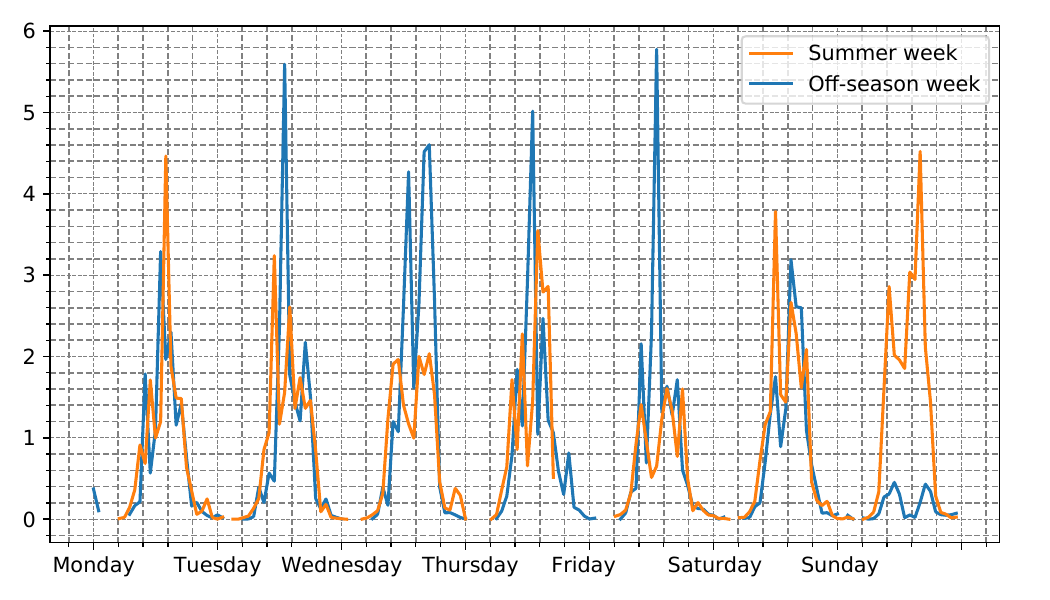}}
\caption{Hourly average number of people at city hall in 2 weeks (Summer week of 2021/08/23 and off-season week of 2022/02/07).}
\label{fig:camera_agosto_versus_fevereiro}
\end{figure}

Another video camera is located in the Smart Lamp Post close to the city hall, pointed to a city leisure park. Figure~\ref{fig:camera_agosto_versus_fevereiro} depicts the hourly average number of people detected, for a span of one week. The idea was to compare two weeks - one from Summer (starting in 2021/08/23) and another off-season (starting in 2022/02/07). The most notorious difference lies on Sundays where, during the Summer, the number of people in the park is higher for the entire day when compared to the Winter season.

\subsubsection{WiFi sniffing}

\begin{figure}[t]
\centerline{\includegraphics[width=9cm]{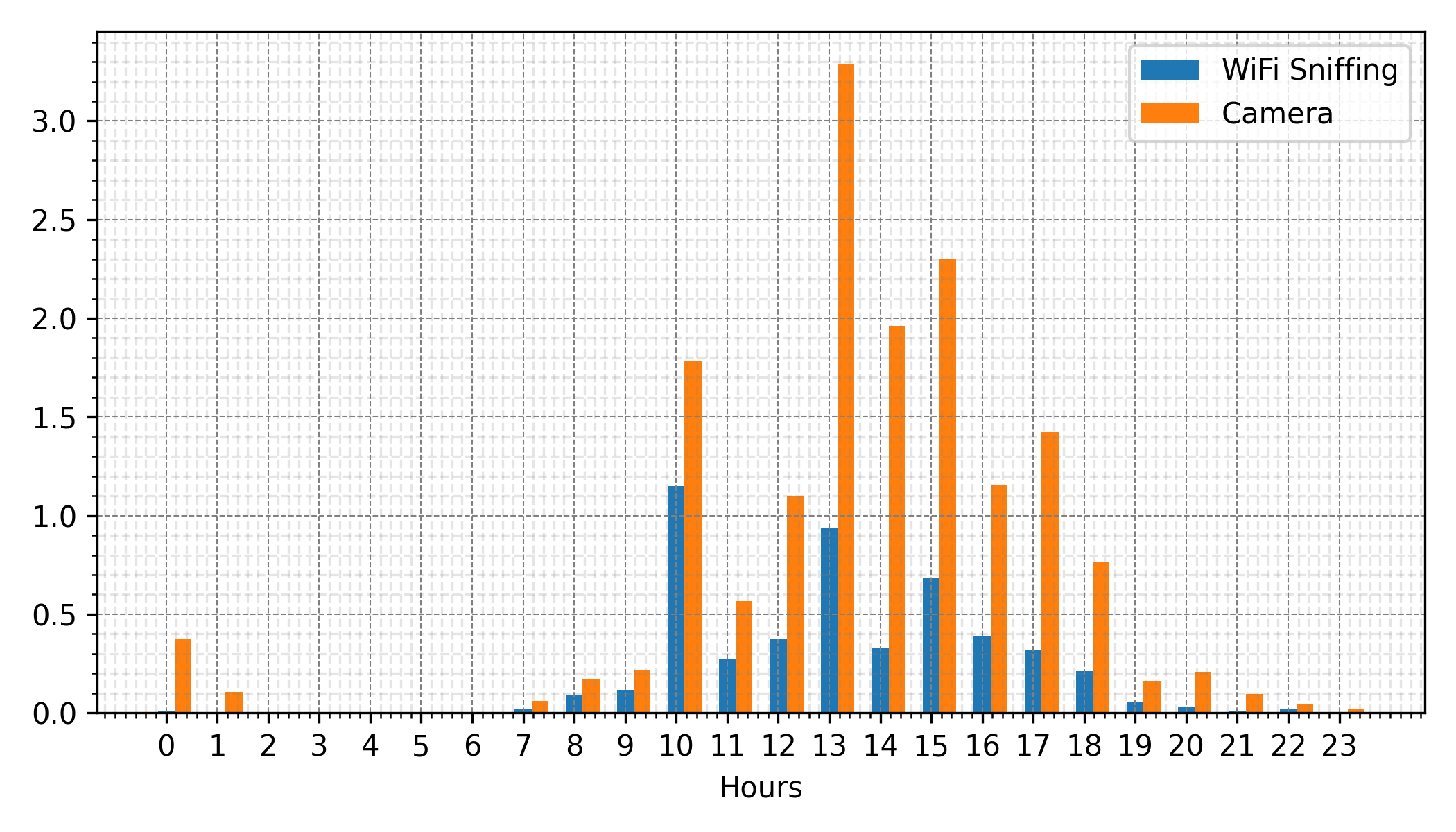}}
\caption{Hourly average of the number of people detected: WiFi sniffing versus camera at city hall (2022/02/07).}
\label{fig:camera_versus_sniffing_p22}
\end{figure}

The WiFi sniffing service uses one of the \ac{rsu}'s wireless network adapter, set in monitor mode, so that it can capture probe requests emitted by the devices that pass near the posts. Assuming that most people carries at least one device with WiFi capabilities, the collected data allows to understand which are the hours when more people are passing by the monitored areas. It is also possible to correlate this data with the detection of the video camera. Figure~\ref{fig:camera_versus_sniffing_p22} plots the results for the WiFi sniffing of devices and video camera detection of pedestrians in the Smart Lamp Post close to the city hall. It is important to note the following differences between these two types of detection: the WiFi sniffing can capture more than one device per person emitting packets trying to join an WiFi \ac{ap}, and it does the detection in an omnidirectional coverage; alternatively, the camera identifies each person, but in a range-defined area. Thus, despite different absolute values, it is still possible to correlate the trend of the amount of people in different hours throughout the day.


\subsection{Performance Results}
\label{sec:results_performance}

The second set of results are related to the performance of the vehicular network and the edge computing.

\subsubsection{Vehicular connectivity}

\begin{figure}[t]
\centerline{\includegraphics[width=9cm]{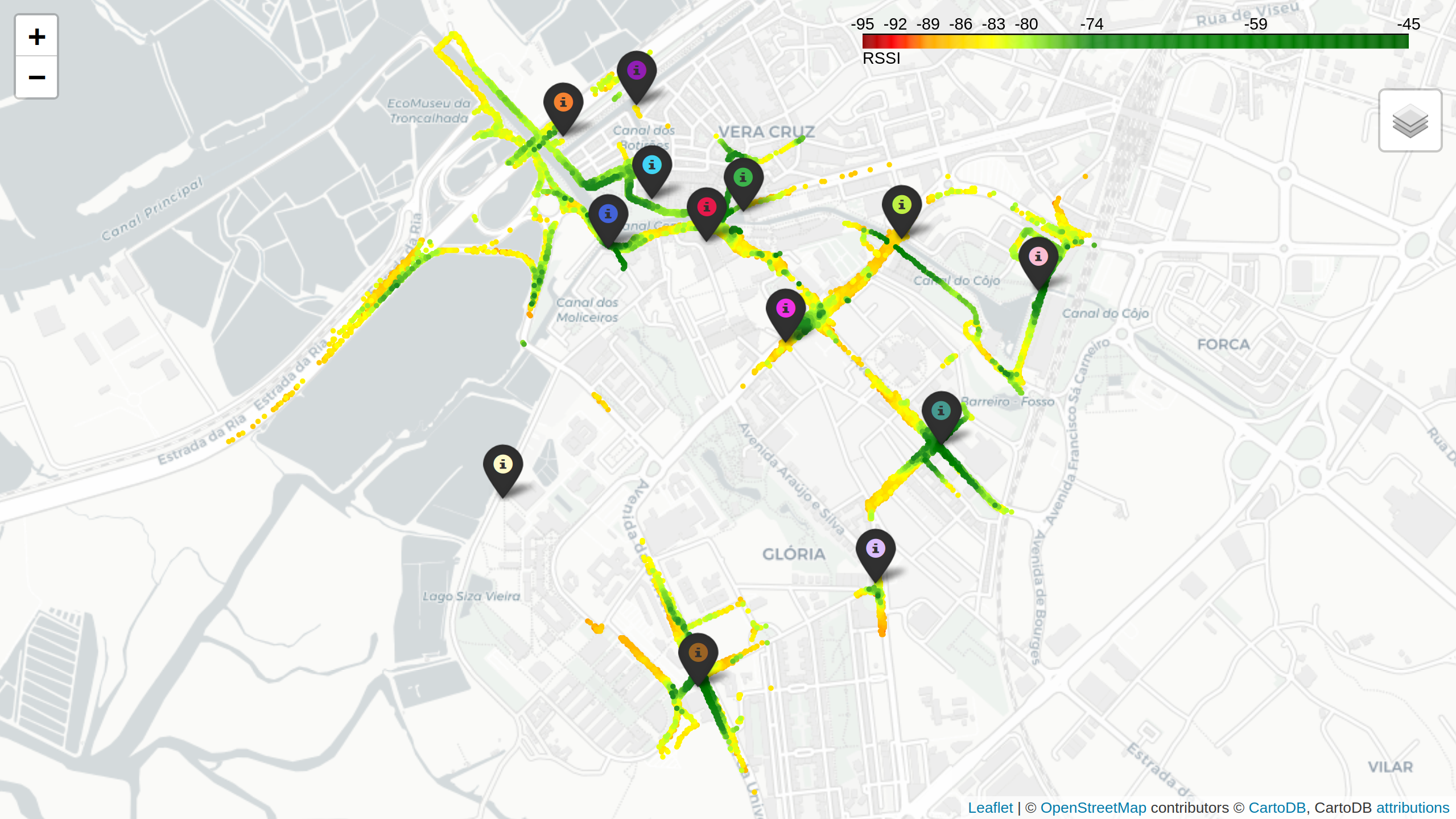}}
\caption{Average \ac{rssi} of \ac{cam} transmissions, as measured on the road-side units.}
\label{fig:global_rssi}
\end{figure}

\begin{figure}[t]
\centerline{\includegraphics[width=9cm]{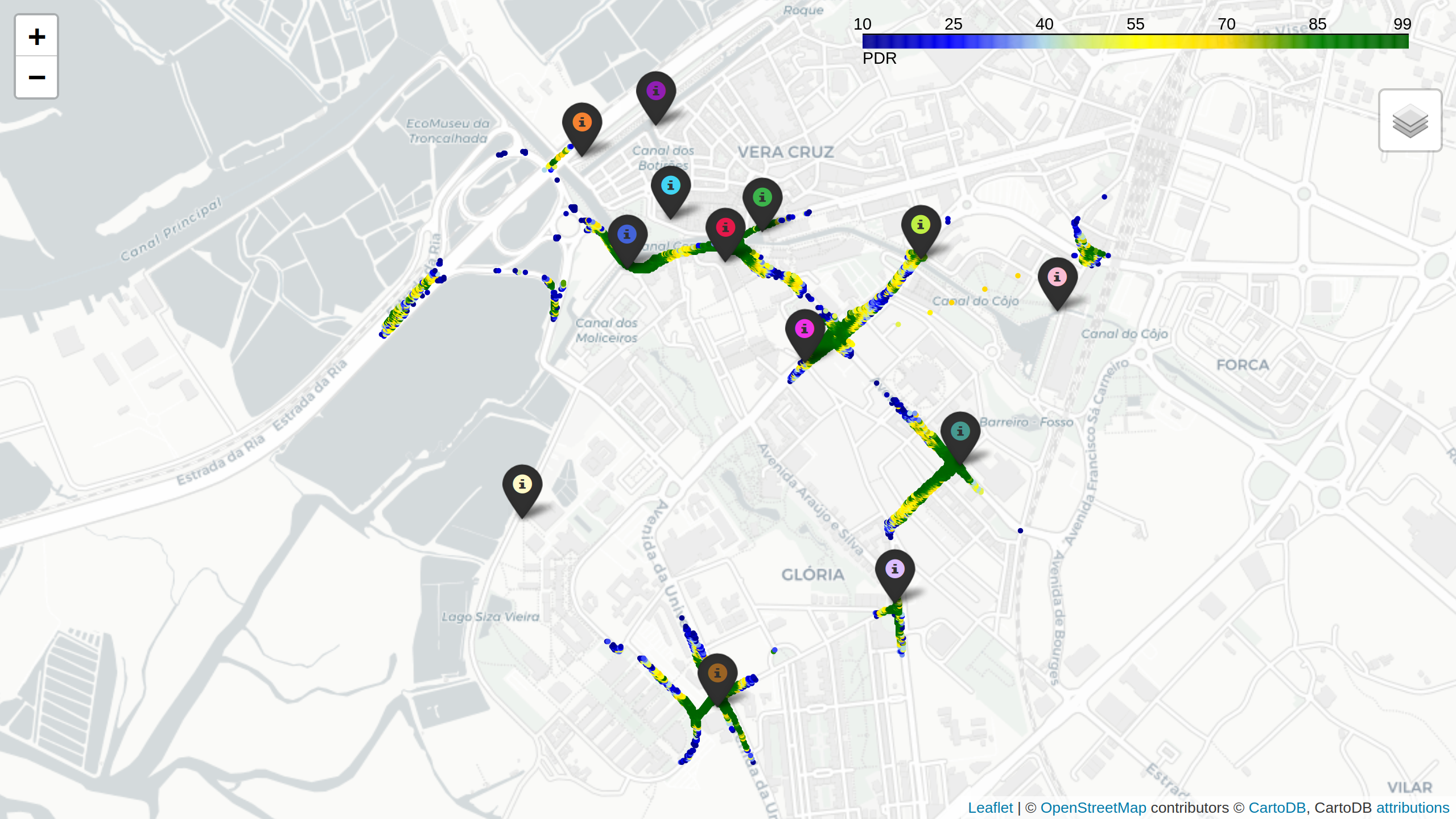}}
\caption{Average \ac{pdr} of \ac{cam} transmissions from vehicles whose transmission rate is known.}
\label{fig:global_pdr}
\end{figure}

In order to obtain initial results on both levels of ITS-G5 coverage and signal quality, maps with results of \ac{rssi} and \ac{pdr} were generated based on \acp{cam} collected during working days (on the week of 2021/04/12). Figure~\ref{fig:global_rssi} plots the average \ac{rssi} measured on the \acp{rsu}. For these results, all received \acp{cam} are considered, including those transmitted by the buses with \ac{atcll}'s \acp{obu}, and conventional vehicles supporting ITS-G5. We can observe a good coverage nearby the \acp{rsu}, in particular in the city center where the density of nodes is higher. Regarding the \ac{pdr} results, the measurements are shown in Figure~\ref{fig:global_pdr}. Here, third-party \acp{cam} were discarded in favour of messages originating only from buses with \ac{atcll} \acp{obu}, since the \ac{cam}'s transmission rate is pre-known. For this reason, the measurements are plotted in a smaller amount of roads. Once again it is possible to observe a high \ac{pdr} nearby the \acp{rsu} and in the city center, with more density of nodes.

\begin{figure}[t]
\centerline{\includegraphics[width=\linewidth]{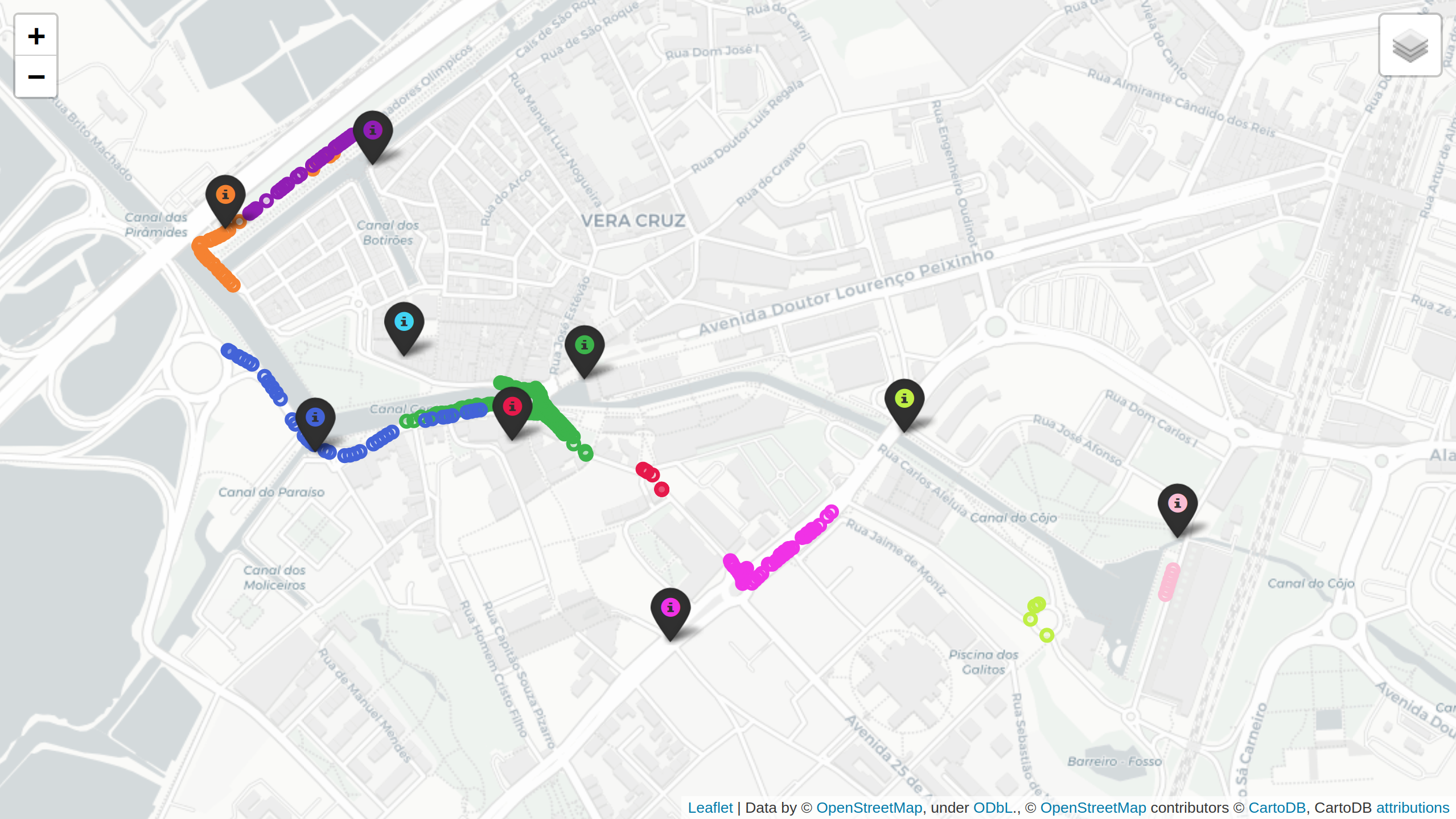}}
\caption{Connectivity tests regarding the vehicular network assessed by a single \ac{obu} in a specific track. The colors represent the connected \ac{rsu} for that location.}
\label{fig:tests_rsu}
\end{figure}

\begin{figure}[t]
     \centering
     \begin{subfigure}[b]{0.48\linewidth}
         \centering
         \includegraphics[width=\textwidth]{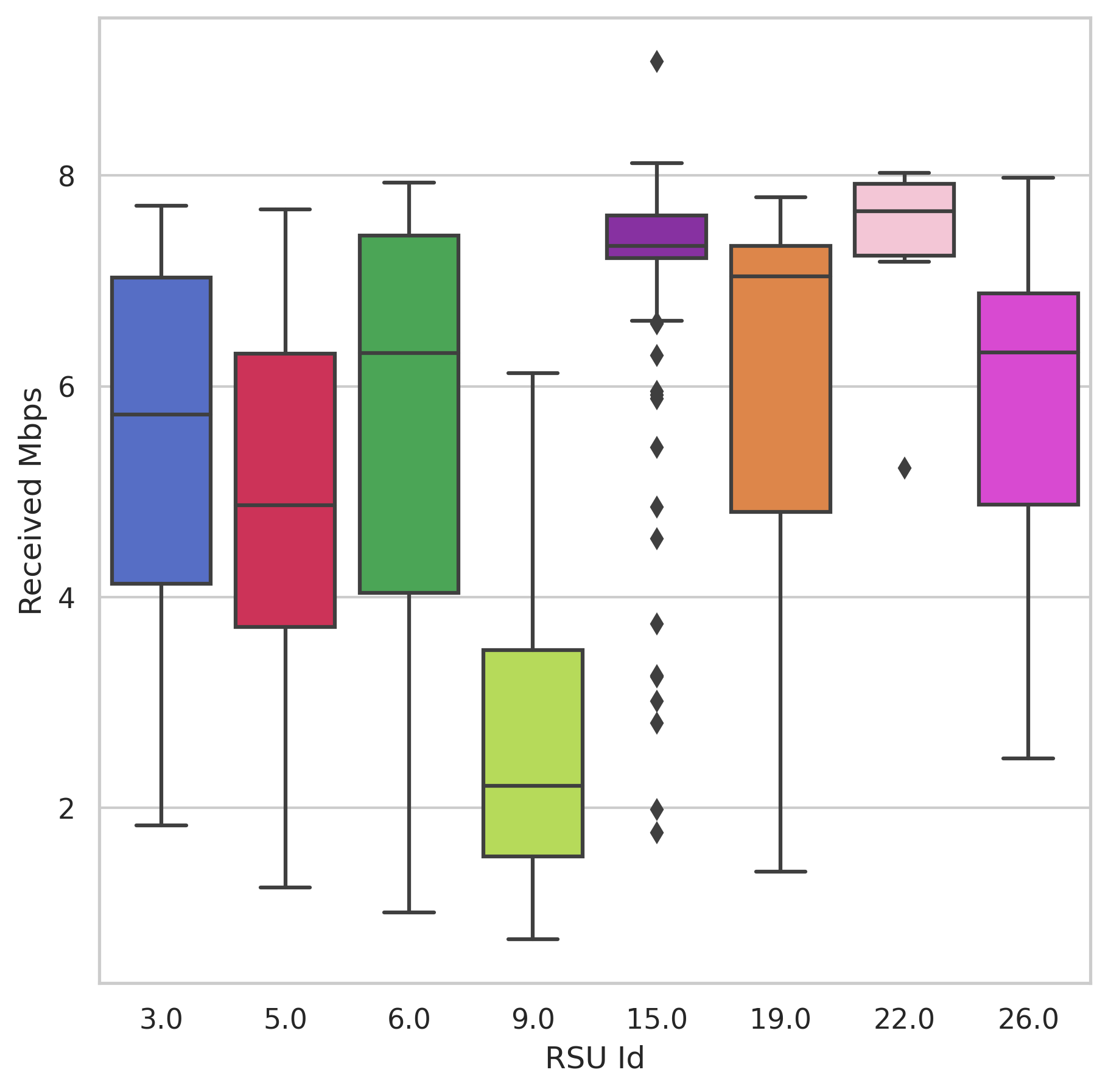}
         \caption{Received Mbps.}
         \label{fig:boxplot_received_mbps}
     \end{subfigure}
     \hfill
     \begin{subfigure}[b]{0.49\linewidth}
         \centering
         \includegraphics[width=\textwidth]{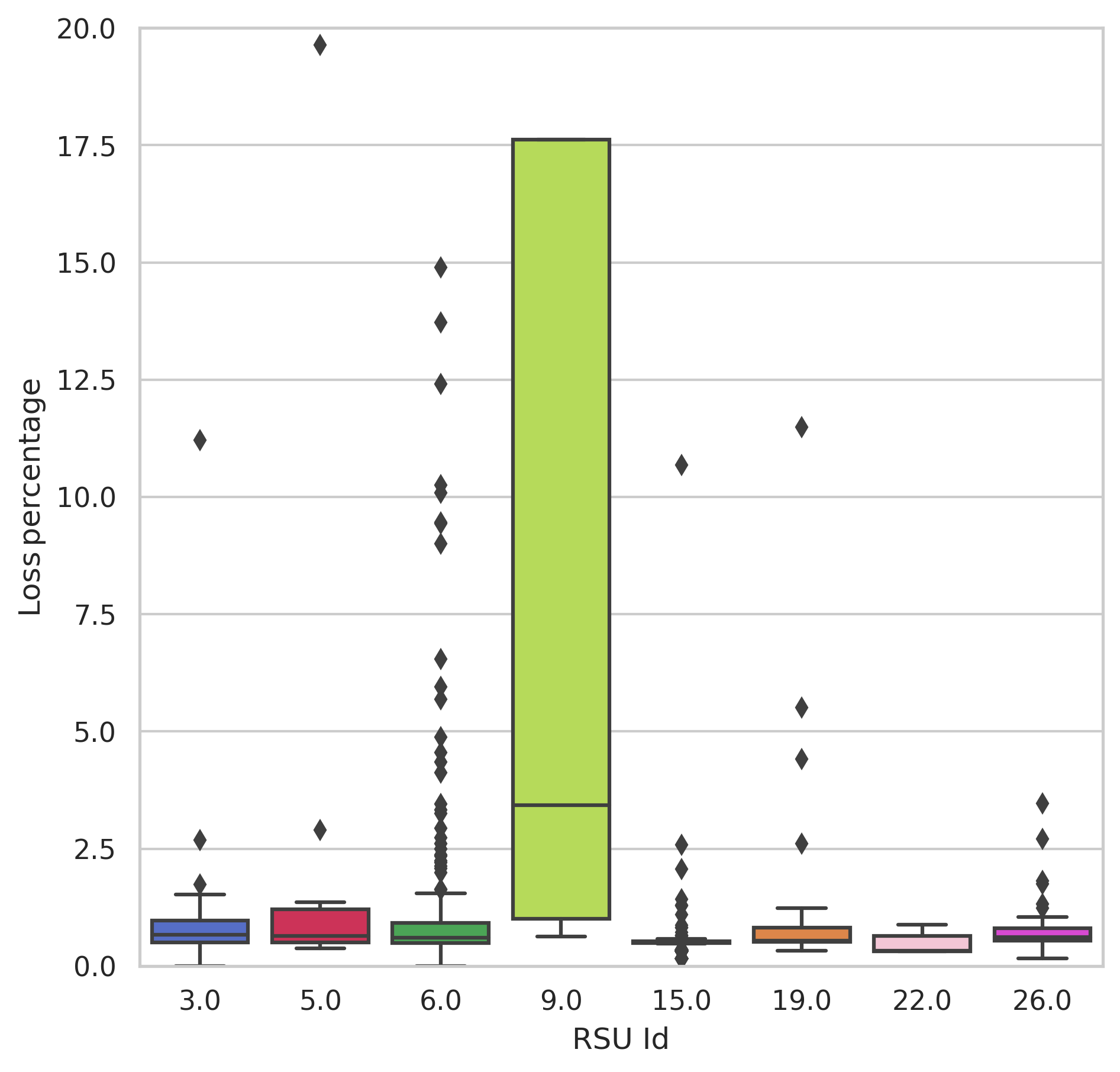}
         \caption{Loss percentage of packets.}
         \label{fig:boxplot_loss_perc}
     \end{subfigure}
        \caption{Network statistics regarding the vehicular network.}
        \label{fig:boxplot}
\end{figure}

In order to evaluate the performance of the initial version of the \ac{sdn}-based mobility network, measurements were performed in the following manner: an iPerf3\footnote{https://iperf.fr} server is running in each \ac{atcll} \ac{obu} of the city buses, and an iPerf3 client deployed in a virtual machine, in the cloud platform, is continuously trying an UDP connection to all \acp{obu}. Upon a successful connection, traffic is generated from the \ac{obu} at packet level, accurately replicating appropriate stochastic processes. The tests were performed with a duration of 1 second per test, with tests being carried out with a frequency of 1~Hz. The \acp{obu} carrying the testing platform circulated through \acp{rsu} numbered P3, P5, P6, P9, P15, P19, P22, and P26, as depicted in Figure~\ref{fig:tests_rsu}. \acp{rsu} P3, P15, P19, and P22 are located in Smart Lamp Posts and \acp{rsu} P5, P6, P9, and P26 are located in wall boxes. Figure~\ref{fig:boxplot_received_mbps} depicts the boxplots of the throughput received by each \ac{rsu}, and Figure~\ref{fig:boxplot_loss_perc} depicts the loss percentage of packets for those transmissions. With such results, we can observe that the \ac{rsu}~P9 receives less data than other \acp{rsu}. This result can be explained by the fact that \ac{rsu}~P9 is positioned under a bridge, providing less coverage than the other wall boxes placed on top of buildings.


\begin{figure}[t]
     \centering
     \begin{subfigure}[b]{\linewidth}
         \centering
         \includegraphics[width=\textwidth]{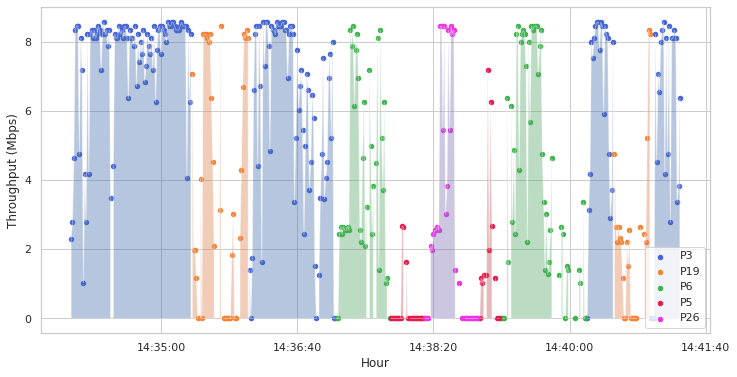}
         \caption{Throughput over time.}
         \label{fig:throughput_time}
     \end{subfigure}
     \hfill
     \begin{subfigure}[b]{\linewidth}
         \centering
         \includegraphics[width=\textwidth]{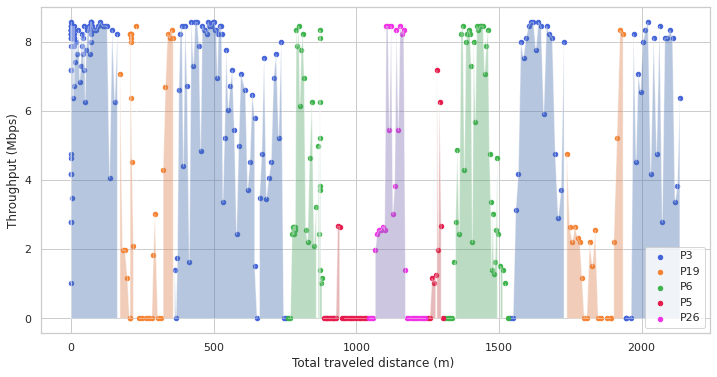}
         \caption{Throughput over distance.}
         \label{fig:throughput_distance}
     \end{subfigure}
        \caption{Collected throughput, received by the OBU, during the connectivity tests over the vehicular network. The different colors represent the connected RSU for that time instant (a), or distance traveled (b).}
        \label{fig:throughput}
\end{figure}

\begin{figure}[t]
\centerline{\includegraphics[width=\linewidth]{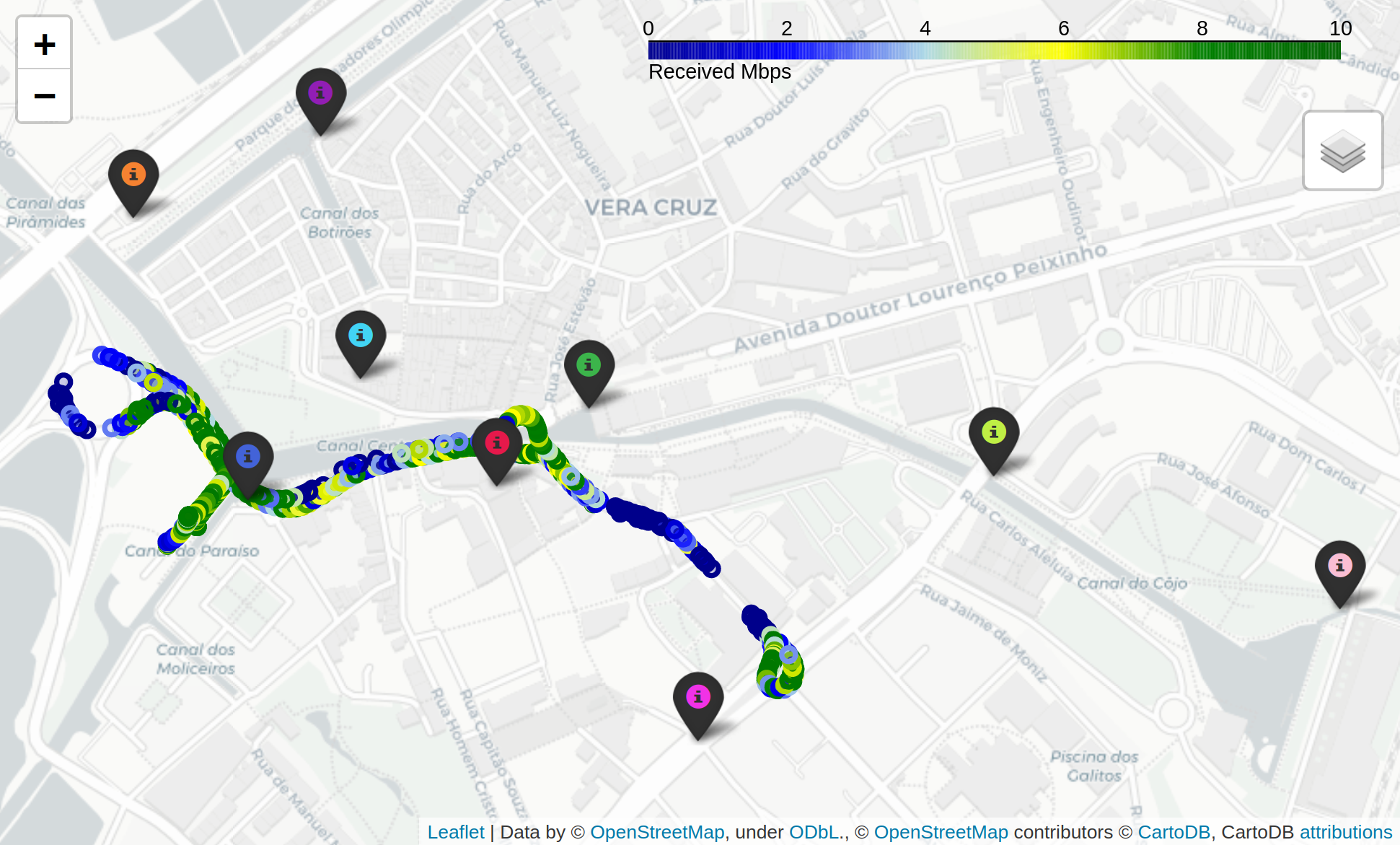}}
\caption{Reception rate (in Mbps) observed in the connectivity tests over the vehicular network, mapped in the city of Aveiro.}
\label{fig:received_mbps_map}
\end{figure}

Figure~\ref{fig:throughput} shows the results of throughput over time and distance traveled in one trip done by a vehicle, passing by several \acp{rsu} (P3, P5, P6, P19, and P26). As corroborated by the results, the current solution of connectivity between \acp{obu} and infrastructure performs well in long periods of connectivity, frequently achieving a peak throughput of, at least, 8~Mbps. These results are also plotted in the map of Figure~\ref{fig:received_mbps_map}, demonstrating city locations where the communication level was sufficient to establish connectivity for the UDP tests, with data on the reception rate (in Mbps) for each position.




\subsubsection{ITS-G5 vs cellular coverage}

\begin{figure}[t]
\centerline{\includegraphics[width=\linewidth]{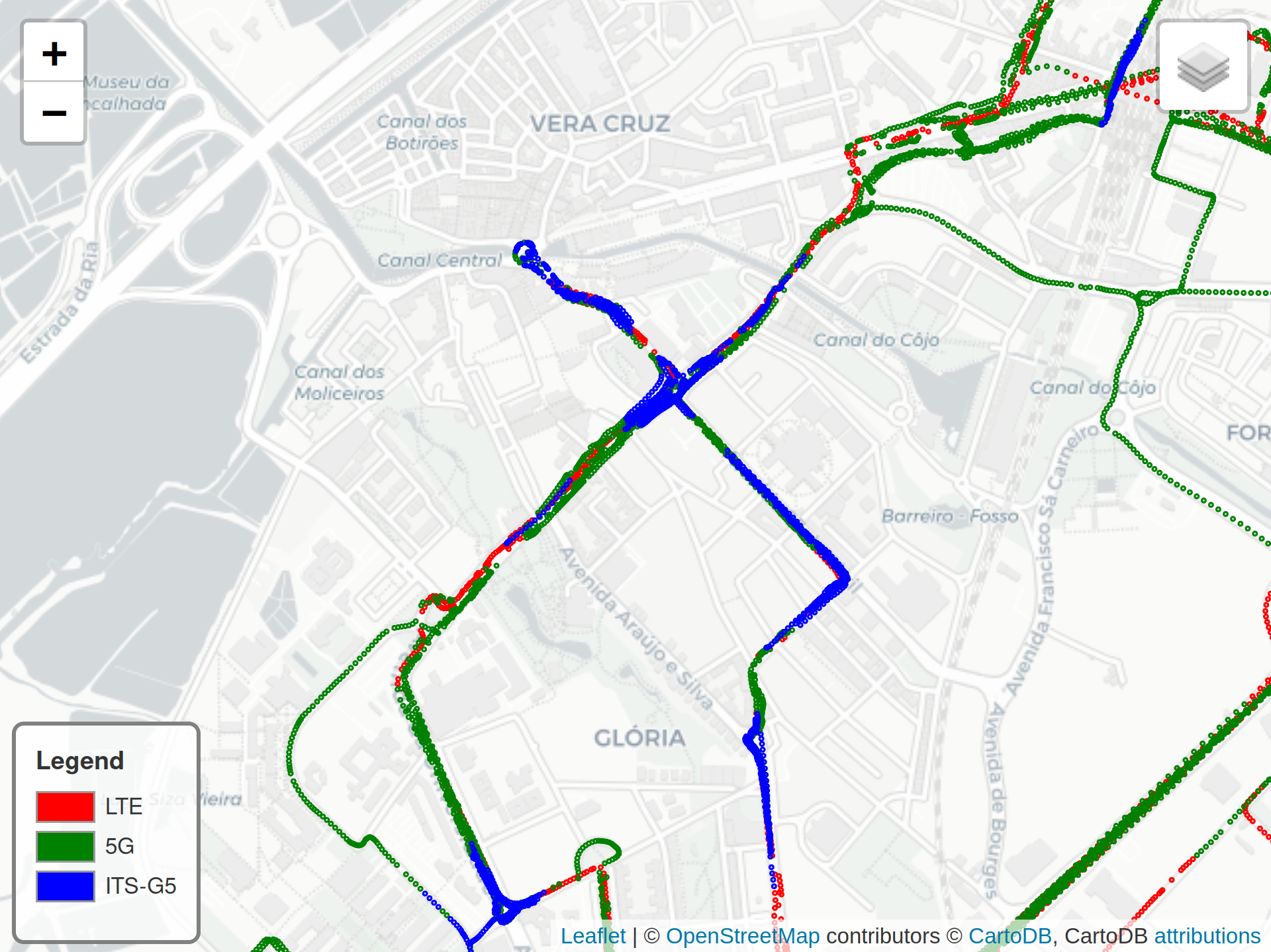}}
\caption{Access technologies for a bus with cellular and \mbox{ITS-G5} connectivity.}
\label{fig:5g_coverage}
\end{figure}

Using a bus with an \ac{obu} supporting both ITS-G5 and cellular (LTE/5G) connectivity, it was possible to plot results of coverage of the different technologies throughout the city. Figure~\ref{fig:5g_coverage} depicts, in different colors, which technology the \ac{obu} used in different areas of the road map of the city. The \ac{obu} was able to use ITS-G5 (in blue) when travelling close to an \ac{rsu}, and changing to 5G (in green) mostly in the other areas. LTE (in red) was used only when the other technologies were not available. The cellular network used in this evaluation was a non-standalone pre-commercial 5G deployment of the mobile network operator Altice MEO (n77~frequency band).

\subsubsection{Edge and Cloud approach}

\begin{figure}[t]
	\centering
	\includegraphics[width=0.85\columnwidth]{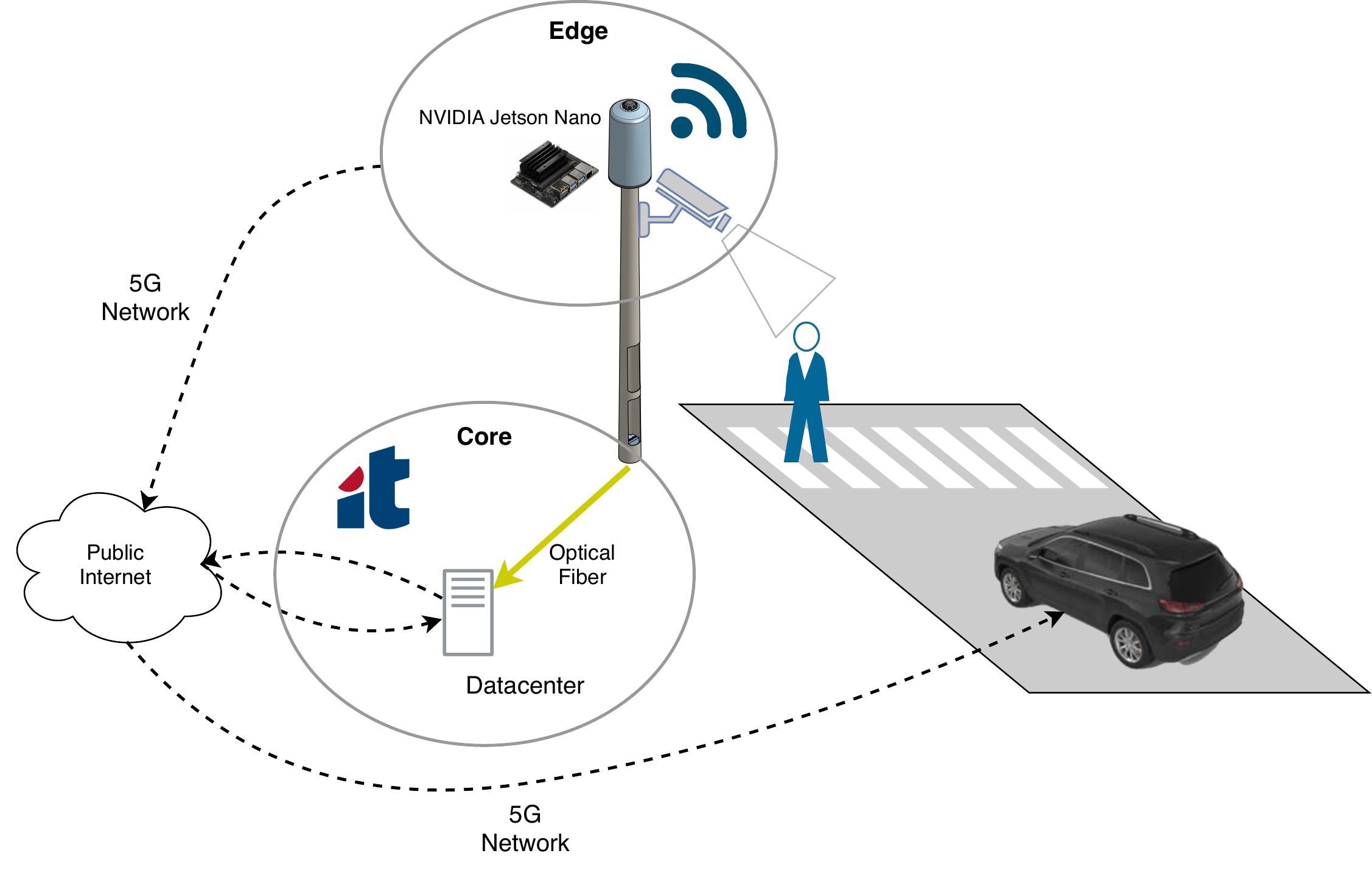}
	\caption{Camera detection architecture diagram \cite{Perna2021}.}
	\label{fig:camera_detection_diagram}
\end{figure}

In order to demonstrate the performance of running services in the edge or in the cloud of the infrastructure, we performed a specific scenario of people detection in the road through video cameras \cite{Perna2021}. Figure~\ref{fig:camera_detection_diagram} depicts the architecture of this scenario: a person (hereby considered as a \ac{vru}) is approaching a crosswalk; next to the \ac{vru} there is a Smart Lamp Post equipped with a video camera, capturing images from that area. The connectivity between the Smart Lamp Post and the Internet is assured by two ways. Following a traditional approach, the Smart Lamp Post is connected to the Internet through dark fiber. Alternatively, to consider some situations where the installation of dark fiber is too expensive, we assume that a 5G network connects the lamp post to the Internet. We evaluated this architecture in a real scenario with real users and vehicles in the area. The 5G technology is the same pre-commercial network as before.

\begin{figure}[t]
\centerline{\includegraphics[width=9cm]{"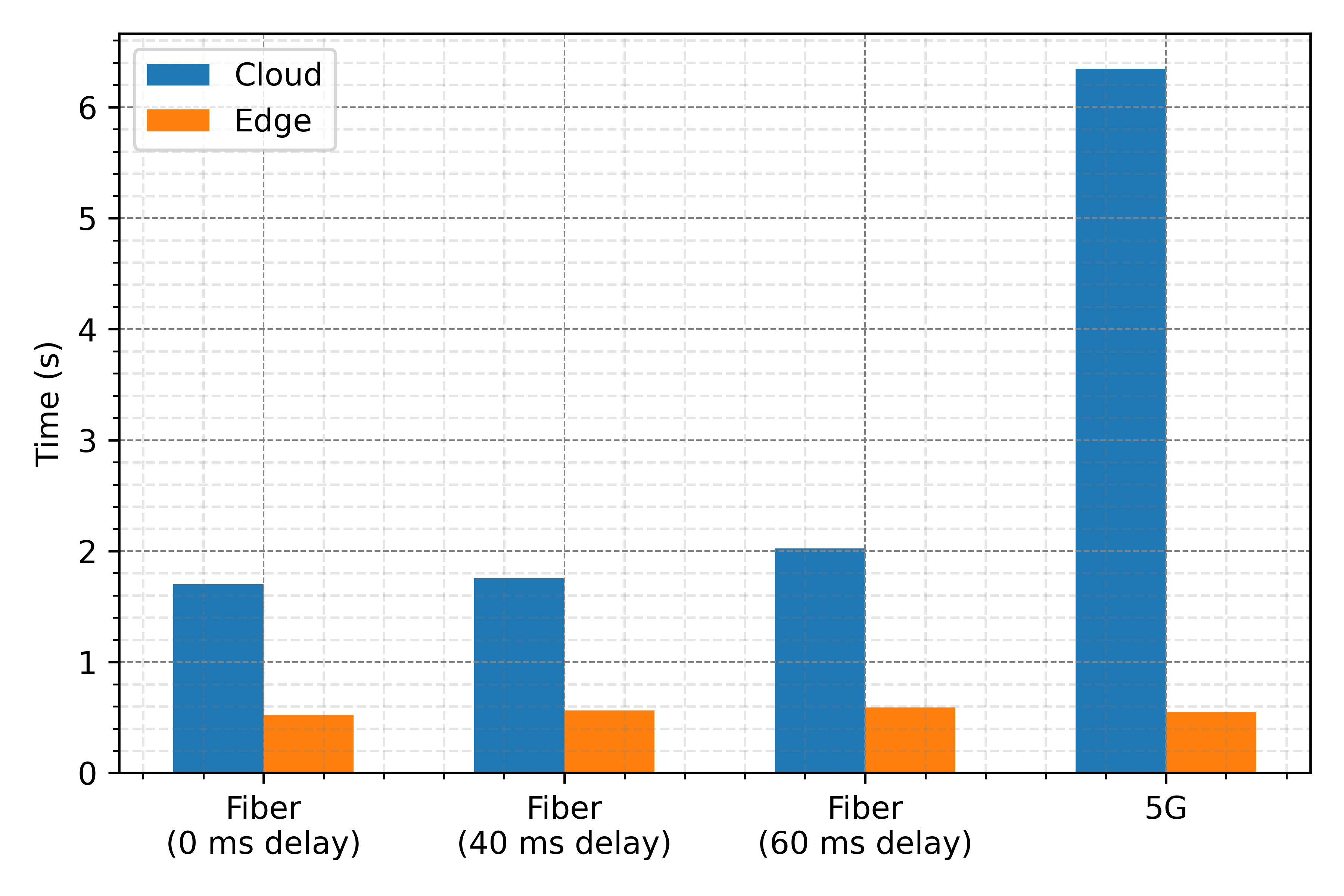"}}
\caption{Detection delay in the edge and cloud for the camera detection module \cite{Perna2021}.}
\label{fig:camera_detection_delay}
\end{figure}

Figure~\ref{fig:camera_detection_delay} compares the detection being processed in the edge and in the cloud with TCP packets. For each bar, the results were obtained by adding the delays of each of the steps that make up the solution flow: the capture delay from the camera becoming available at the Nvidia Jetson; the detection module delay; and also the communication delay between the edge and the cloud. The difference between the results on the edge and the cloud is noticeable, with both the processing and the communication times showing up a strong reduction in the edge-based variant. In this edge approach, only a notification with the detection results (with a length of 41 data bytes) is sent to the cloud. In this case, comparing the performance of the 5G network with the usage of fiber, these delays are now very close. This result is relevant if the 5G option is used preferably due to the stronger flexibility on the positioning of the video camera throughout the city.

\section{Use Cases}
\label{sec:use_cases}

To demonstrate the potential of the communication, computation and sensing platform presented in the ATCLL, we demonstrate a set of different use cases, targeting different verticals ranging from mobility characterization up to collision avoidance and safety of VRUs, including also the assessment of new communication technologies (mmWave) and communication paradigms (such as Informaction Centric Networks - ICNs).



\subsection*{Use-case 1: Mobility characterization} 

Managing traffic can be a difficult task. Traffic jams and parking difficulties are some of the problems that citizens can face in an urban scenario. To help city managers to improve urban mobility, the visualization of vehicle data on the map is critical.

\begin{figure}[t]
     \centering
     \begin{subfigure}[b]{\linewidth}
         \centering
         \includegraphics[width=0.8\textwidth]{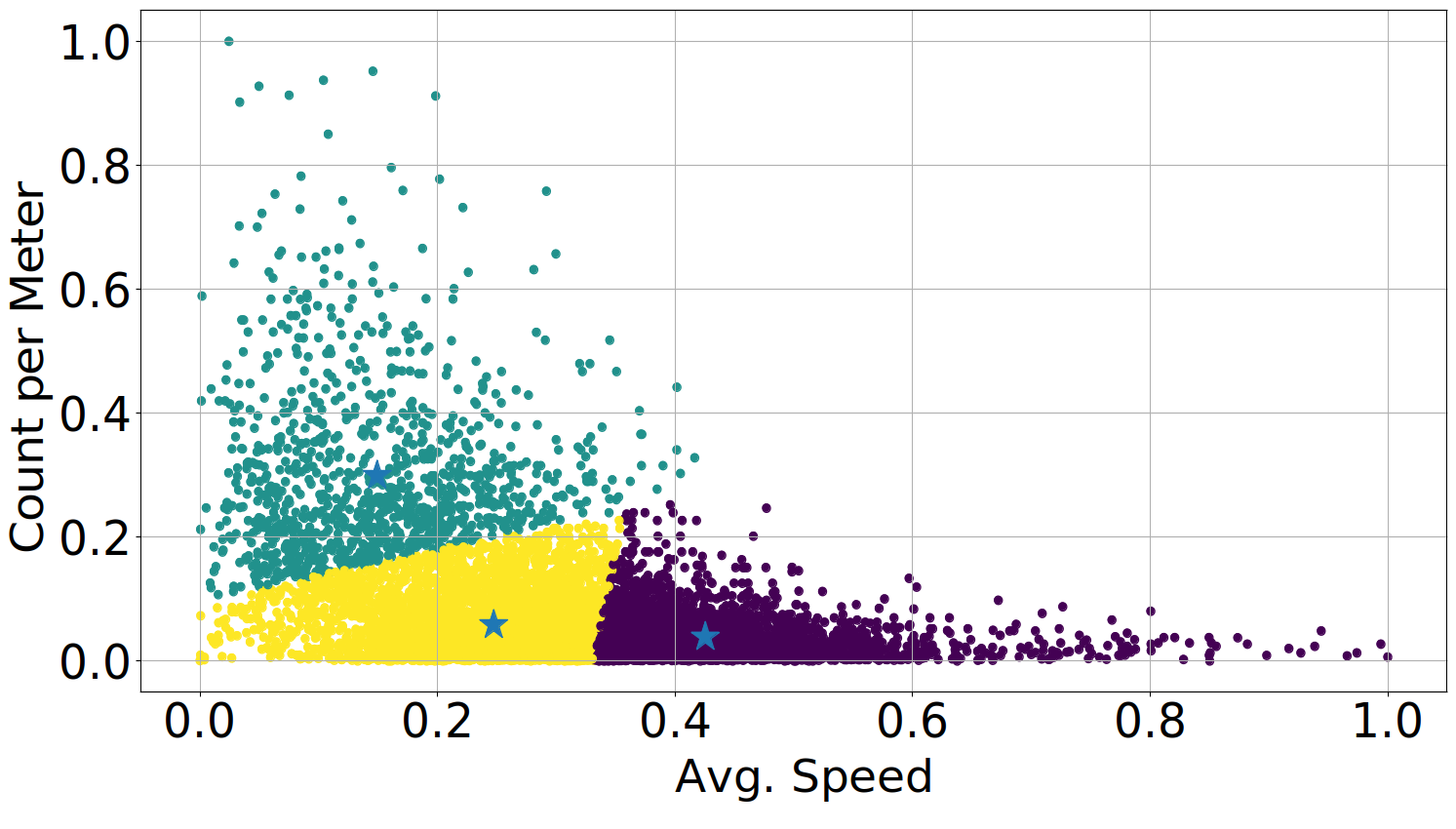}
         \caption{Clustering results regarding the vehicle's speed.}
         \label{fig:clustering_results}
     \end{subfigure}
     \hfill
     \begin{subfigure}[b]{\linewidth}
         \centering
         \includegraphics[width=\textwidth]{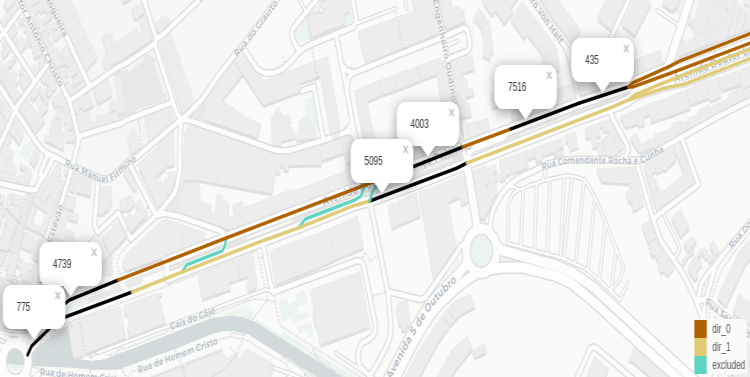}
         \caption{Identification of congestion situations.}
         \label{fig:high_congestion}
     \end{subfigure}
        \caption{Traffic congestion study.}
        \label{fig:traffic_congestion}
\end{figure}

To study traffic congestion, we created a model based on the speed and the count of vehicles per meter~\cite{traffic_congestion}. This information is given by the CAM messages information from the connected buses. Using clustering techniques we detect different levels of congestion as depicted in Figure~\ref{fig:clustering_results}. The cluster with a higher level of congestion has the highest count per meter values and lower speeds. We mapped this information into road segments, as shown in Figure~\ref{fig:high_congestion}. The highest level congestion zones are the identified segments. Thus, it is possible to understand traffic congestion, and identify temporal and spatial patterns. 

\begin{figure}[t]
\includegraphics[width=\linewidth]{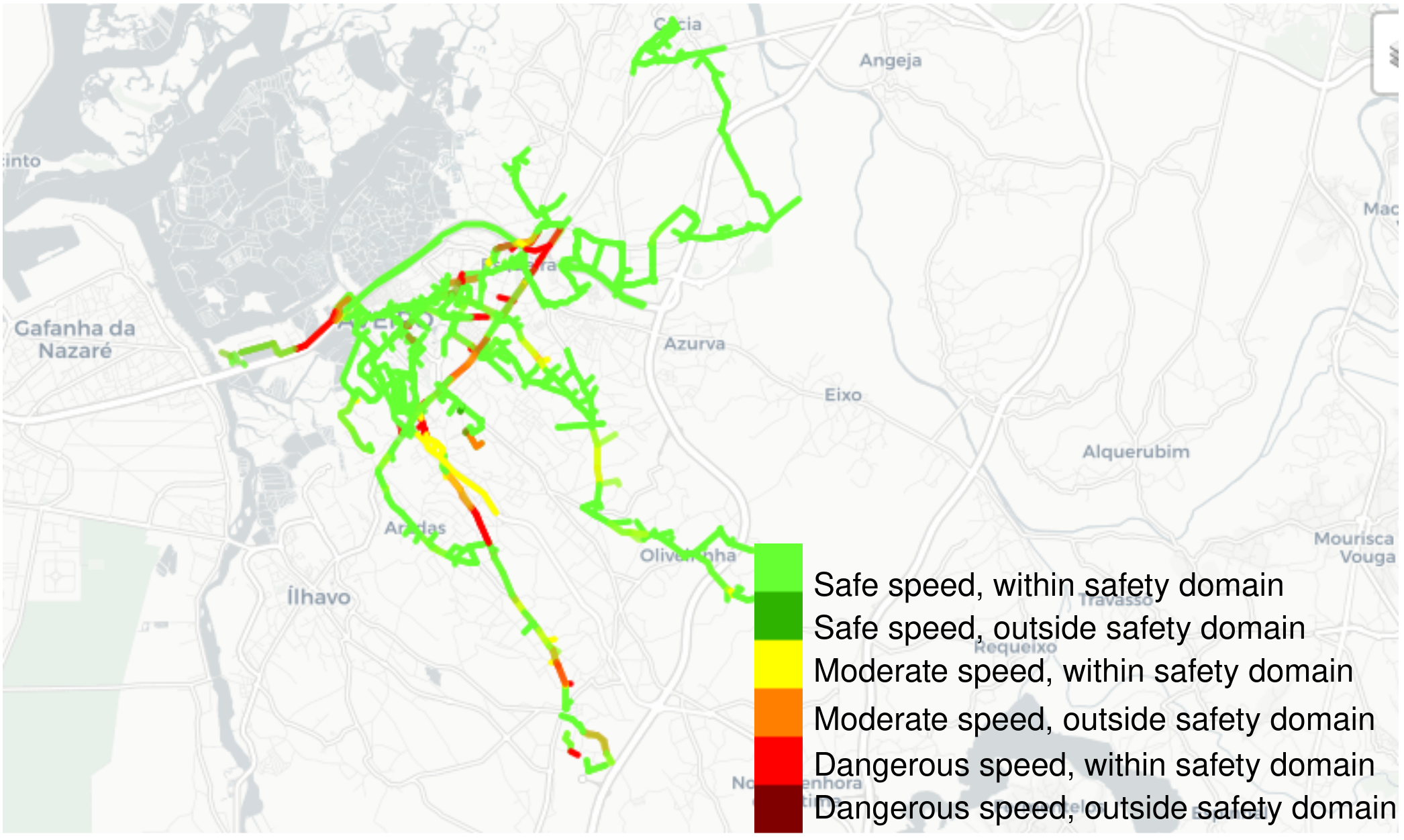} 
\caption{Driving behavior study \cite{dbehavior}.}
\label{fig:driving_behavior}
\end{figure}

Unsafe driving behavior can lead to road accidents. The identification of problematic road segments and zones becomes essential to help city managers to improve road safety. Through the mobility data, we proposed a model to classify the driving behavior~\cite{dbehavior}, taking into account the speed of the vehicle, the speed limit, the acceleration, and the road friction. Figure~\ref{fig:driving_behavior} contains the six created categories: the speed indicates if the legal speed limits are respected; regarding safety, stating "within safety domain" means that no sudden changes in the acceleration were measured, in opposition to "outside the safety domain", meaning sudden accelerations or decelerations were performed.

\subsection*{Use-case 2: Collision avoidance with vulnerable road users}
\begin{figure}[t]
    \centering
    \includegraphics[width=\columnwidth]{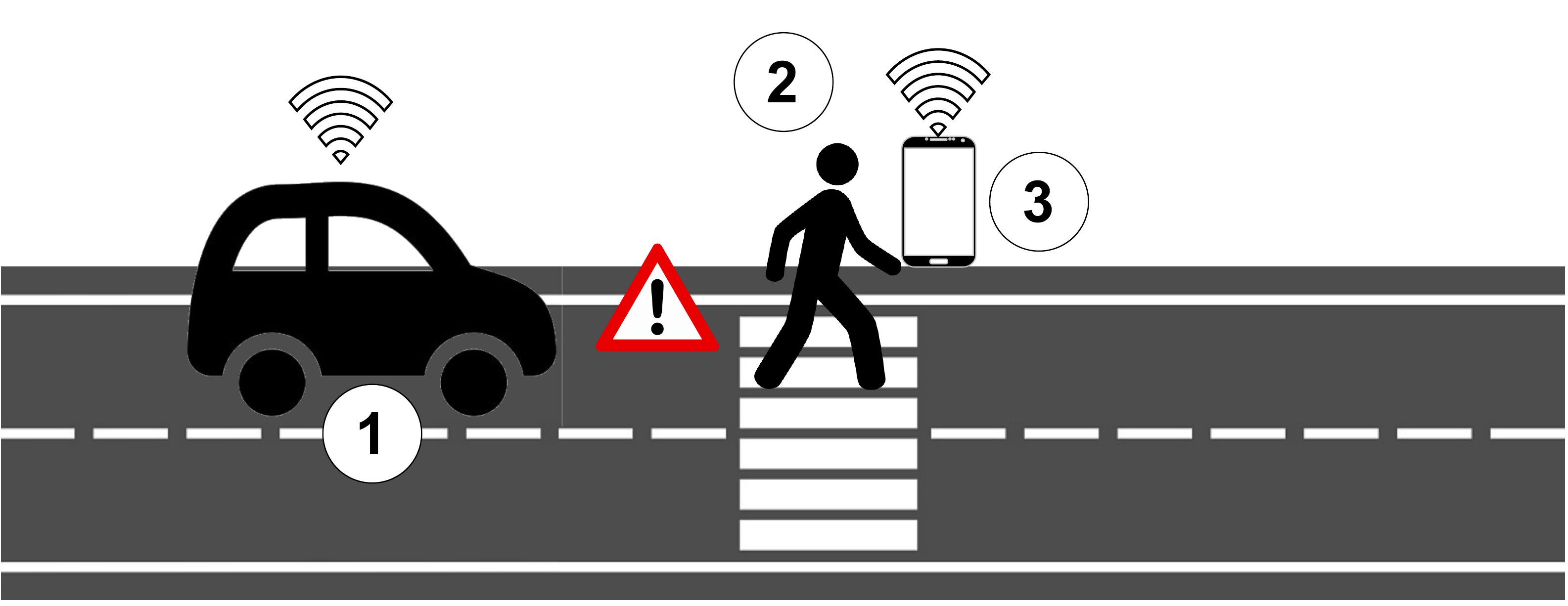} 
    \caption{A vehicle (1) and a pedestrian (2) are warned about a potential collision. The pedestrian has an \ac{its-s}-capable device, e.g. a smartphone (3) \cite{PTeixeira}.}
    \label{fig:problem-use_cases_goals}
\end{figure}

\Aclp{vru} are particularly susceptible to accidents with vehicles. To minimize potential collisions, a system capable of detecting potential accidents was developed~\cite{PTeixeira}. Figure~\ref{fig:problem-use_cases_goals} shows the scenario where a vehicle and a pedestrian are in a situation of a potential collision. The pedestrian (\ac{vru}) is an \ac{its-s} through the smartphone and can be notified about an imminent accident. The vehicle is also notified through its \ac{obu}. Thus, they can adopt strategies to solve the situation (either slowing down, braking, changing direction, among other actions). To improve accuracy, the system was developed considering the data fusion from several edge nodes — traffic radar, camera detection, vehicles, and the \acp{vru} own smartphones.

\begin{figure}[t]
    \centering
    \includegraphics[width=\linewidth]{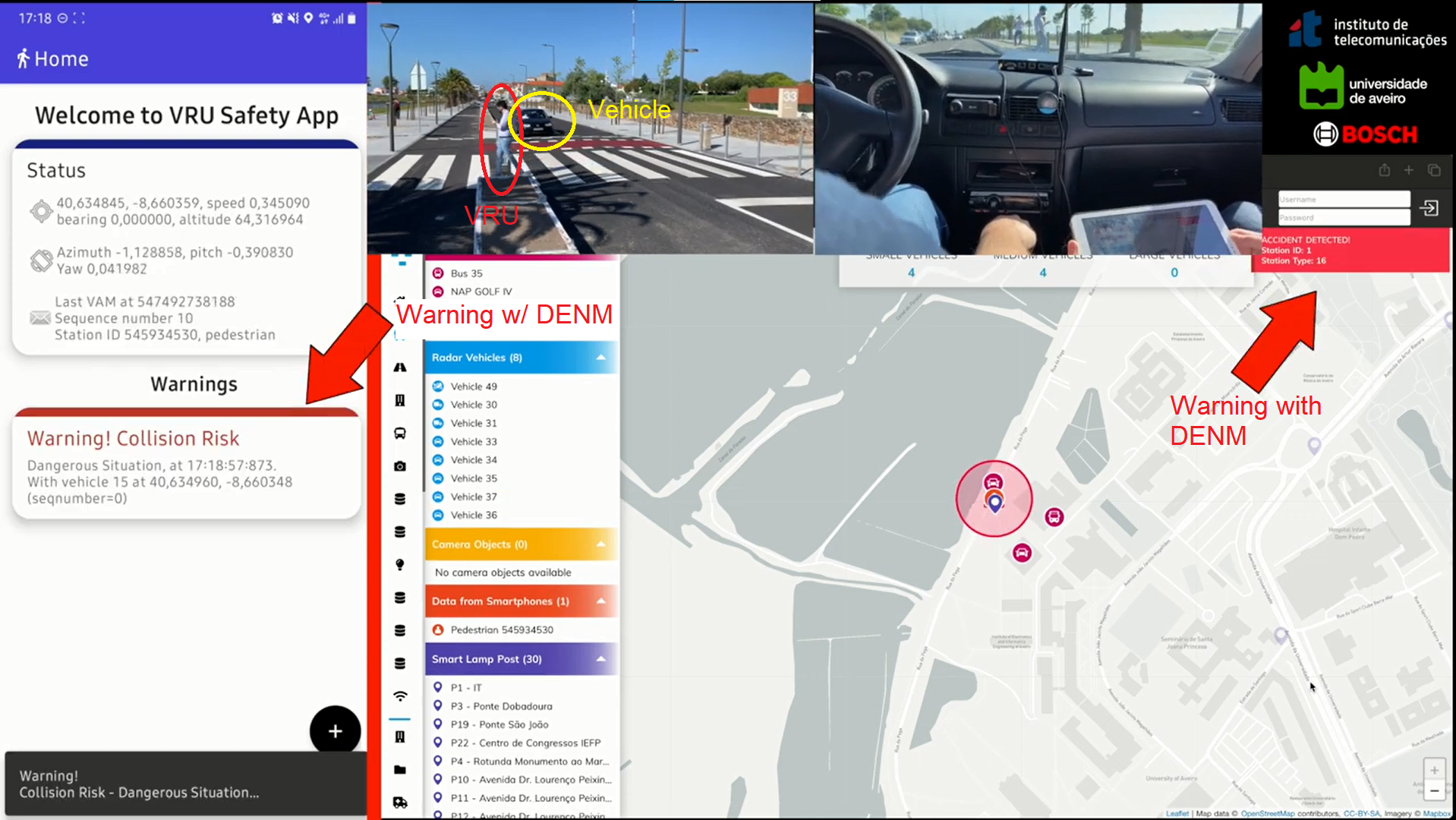}
    \caption{VRU test execution and results~\cite{PTeixeira}.}
    \label{fig:evaluation-accuracy-test1-results}
\end{figure}

Figure~\ref{fig:evaluation-accuracy-test1-results} presents results of a real test, showing the overall end-to-end operation of the system, and that the system is capable of detecting the potential collisions, and warn both the \ac{vru} and vehicle, through the \ac{vru} \ac{its-s} smartphone application and the \ac{atcll} dashboard, respectively.



\subsection*{Use-case 3: Emergency Vehicles communication}
\begin{figure}[t]
\centering
\includegraphics[width=0.85\linewidth]{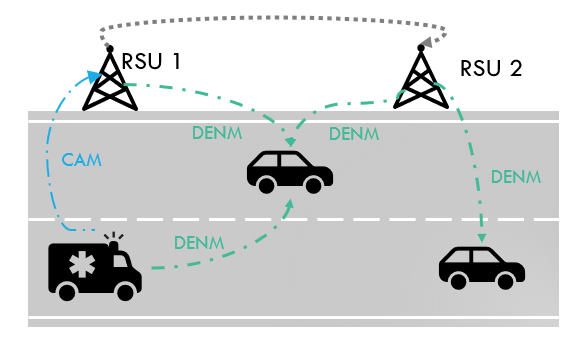} 
\caption{Message dissemination when an EV approaches.}
\label{fig:emergency_use_case_final}
\end{figure}

Nowadays, when an \ac{ev} responds to an emergency, it alerts the road user of its presence through sound and light. However, road users can only detect its presence when the \ac{ev} is nearby, making it difficult to clear the road as quickly and safely as needed without interfering with the \ac{ev} course. To minimize the delay in the \ac{ev} route, a system that automatically alerts road users near an \ac{ev} through the dissemination of \acp{denm} was developed. Figure~\ref{fig:emergency_use_case_final} shows a set of messages sent to alert road users near an \ac{ev}.

\begin{figure}[t]
\centering
\includegraphics[width=\linewidth]{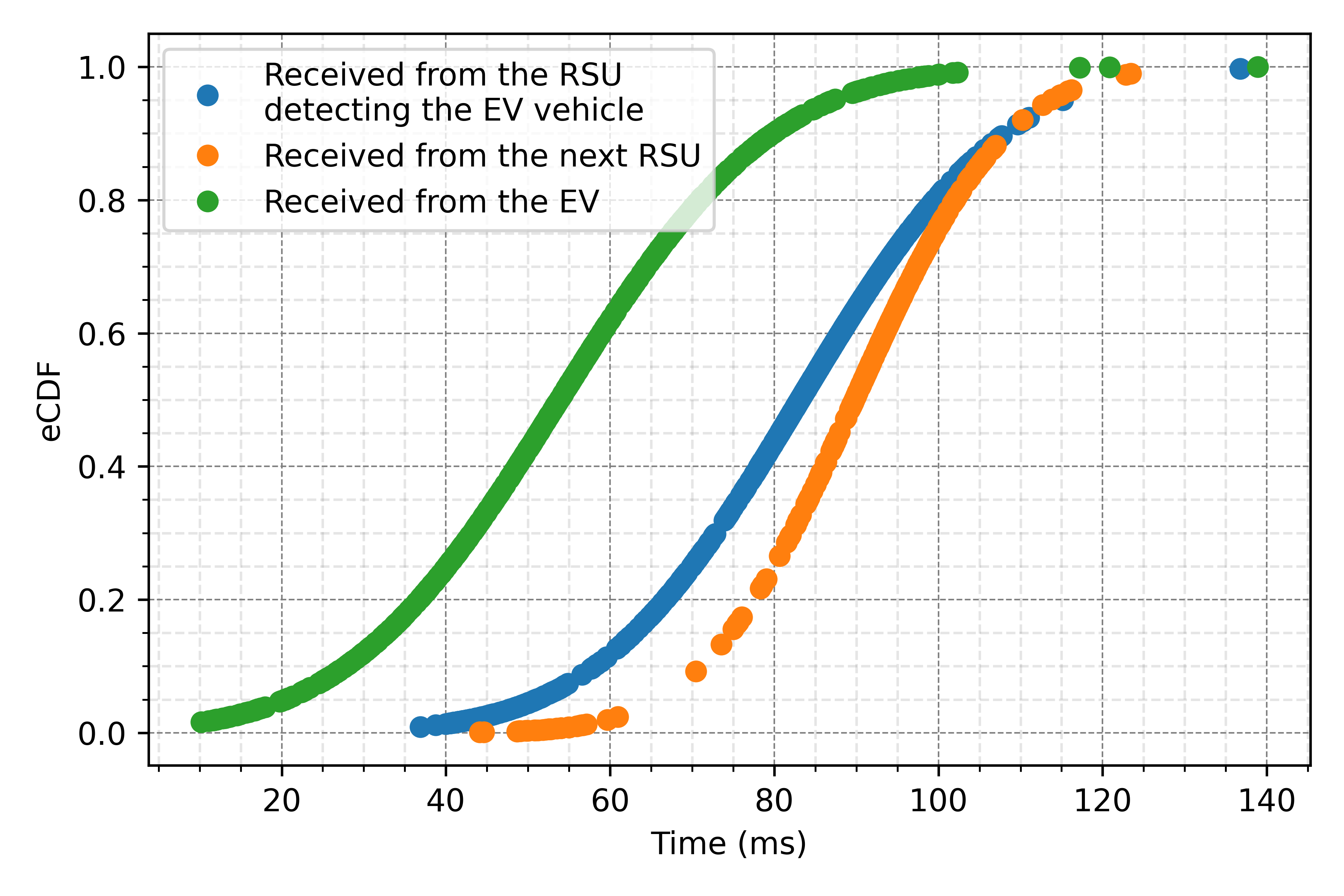} 
\caption{Message dissemination latencies~\cite{Andreia}.}
\label{fig:emergency_latency}
\end{figure}

Tests in the ATCLL were performed to evaluate the system designed for this use case. One of the metrics evaluated in those tests was the delay time between the detection of an \ac{ev} and the reception of a \ac{denm} by other road users. Figure~\ref{fig:emergency_latency} represents the empirical cumulative distribution function of the system's total delay time for three different \ac{denm} transmitters: the \ac{ev}, the \ac{rsu} detecting the \ac{ev} (e.g. \ac{rsu} 1 of Figure~\ref{fig:emergency_use_case_final}), and the upcoming \ac{rsu} (e.g. \ac{rsu} 2 of Figure~\ref{fig:emergency_use_case_final}).

When the \ac{ev} sends the \ac{denm}, it obtains the lowest delay of the three cases; however, it also obtains the smallest delivery range. On the other hand, for the case where the \ac{rsu} is responsible for the \ac{denm} transmission, both the range and the delay time increase. In the worst case, i.e., when it is the upcoming \ac{rsu} to send the \ac{denm}, it is received with a maximum delay of 108~ms (in 90\% of the cases) and a median time of 89~ms.

In addition to warning messages sent to the road users, the \ac{ev} can also receive real-time information about highly congested areas from the infrastructure. This will allow the \aclp{ev} to avoid congested areas, improving its response time.

In the future, we are hoping to expand this emergency use-case from a vehicle's communication scale to a full-city-scale, where \ac{tsn} nodes~\cite{Seol2021} are also expected to allow latency- and jitter-strict emergency communications to happen, bringing the time concept towards our infrastructure.

\subsection*{Use-case 4: Industrial Internet-of-Things to cloud continuum}

\begin{figure}[t]
\centerline{\includegraphics[width=\linewidth]{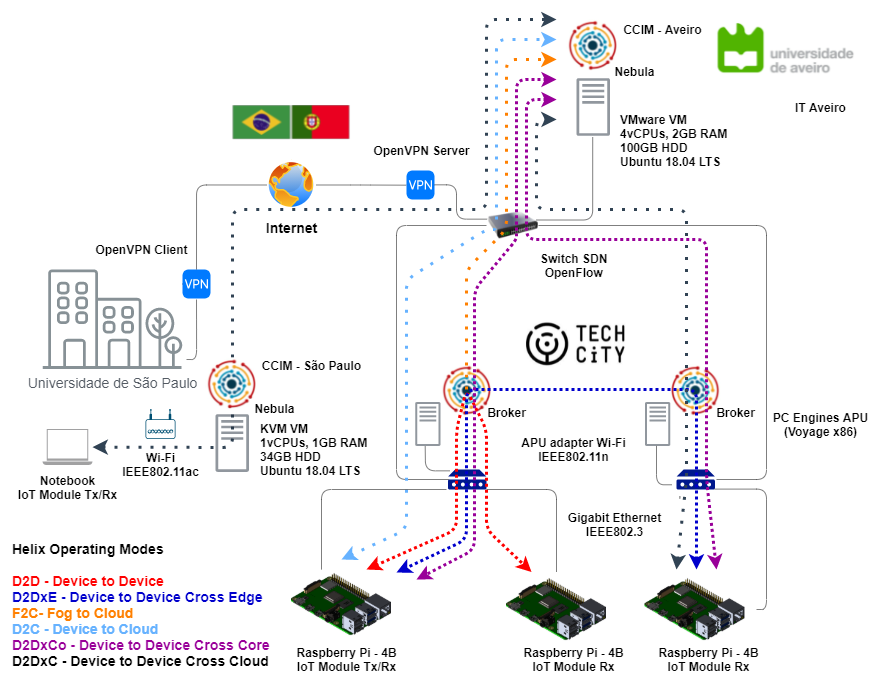}}
\caption{Infrastructure used to evaluate the Helix Multi-layered platform at the ATCLL~\cite{Cabrini21}.}
\label{fig:Helix}
\end{figure}

The \ac{iiot} is one of the most demanding \ac{iot} applications. The integration of industries in the context of smart cities and other smart environments, allied with new communication technologies such as 5G, brings a new horizon of possibilities and new requirements. To attend this type of application, we studied the integration of the Helix Multi-layered \ac{iot} platform and its operating modes into the \ac{atcll}, showcasing a real-world deployment
~\cite{Cabrini21}. The Helix Multi-layered platform was deployed and evaluated in this city platform to identify its potential to support services and applications that require operation in an edge environment, as depicted in Figure~\ref{fig:Helix}.

\subsection*{Use-case 5: mmWave backhaul research}

Millimeter-wave (mmWave) could be seen as an attractive option for the \ac{iab}. Due to the capability to support \mbox{multi-gigabit-per-second} data rates, the technology allows the reduction of the deployment expenses of fiber optics, with the introduction of dense cell deployments. However, the use of higher frequencies (e.g., \mbox{V-E Band}) is not idle by chance, mmWave communications suffer from strong path loss, and rain atmospheric propagation absorption, making them only suitable for communication in short and \ac{los} communications. The advances with the introduction of small antenna arrays allow the use of high dimensional and directionally steerable beams, that are formed by sophisticated beamforming techniques. Such techniques allow to ameliorate the propagation characteristics, but with a high cost in complexity.

\begin{figure}[t]
\centerline{\includegraphics[width=\linewidth]{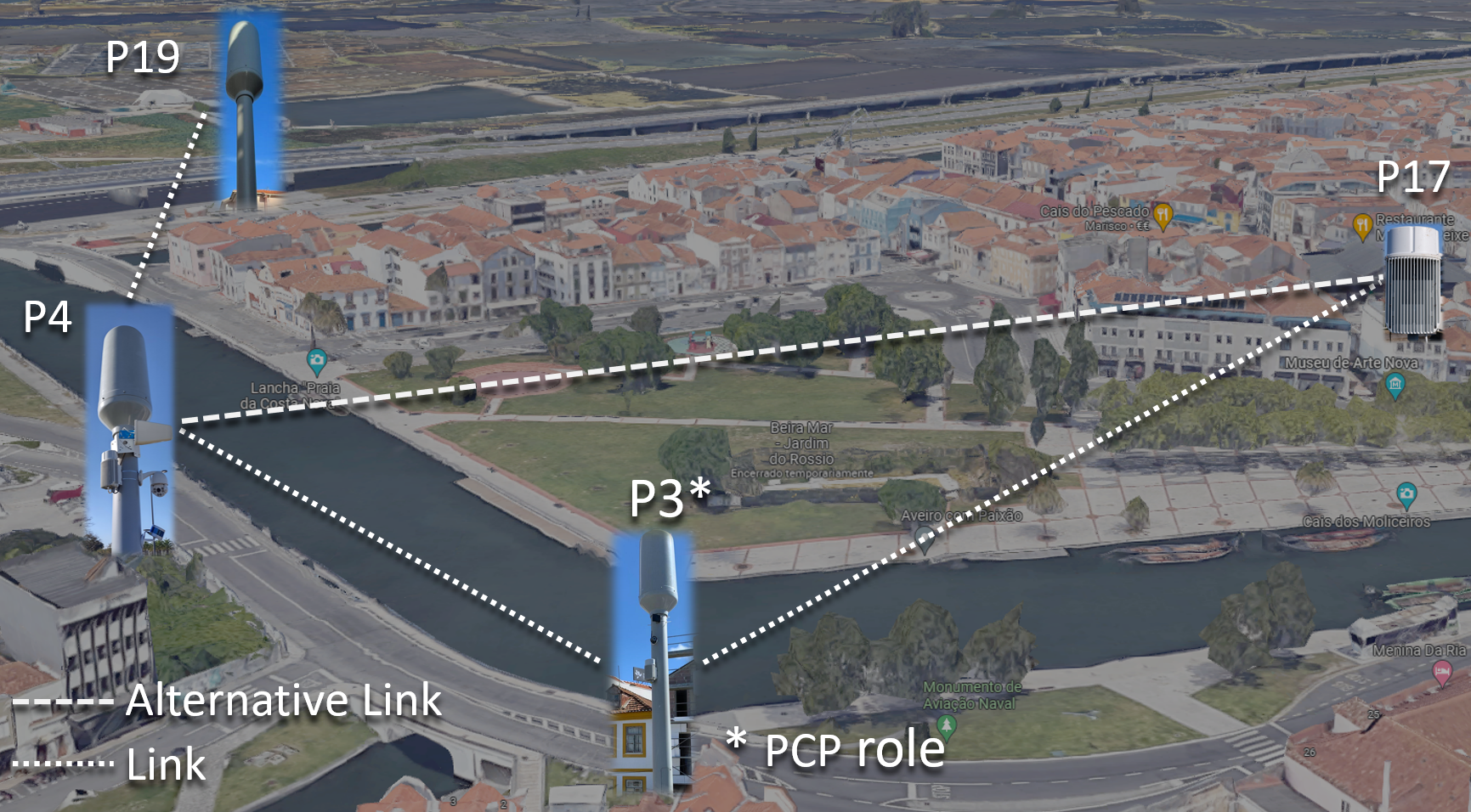}}
\caption{Millimeter-wave backhaul network deployed in the ATCLL, also as a detail of Figure~\ref{fig:ATCLL_map}.}
\label{fig:mmwave}
\end{figure}

The challenges of the mmWave could be found at various levels of the protocol stack, and usually differ from other wireless technologies, because of their need for \ac{los}. Thus, in this context, the \ac{atcll} presents an ideal research support platform to evaluate and propose new mechanisms capable to deal with the lossy nature introduced by the mmWave communications (e.g., link quality assessment, rate adaptation and bufferbloat in the transport layers). In this context, the mmWave deployed in the ATCLL (Figure~\ref{fig:mmwave}) has contributed to the validation of several research proposals in collaboration with other research groups and companies. Regarding the modeling of blockage in the simulation environments, the work developed in~\cite{mmwave_alexandre} presents a new blockage model that allows the modeling of blockage scenarios in the IEEE802.11ad standard. In this research proposal, the city testbed was used to compare a real scenario with the blockage model, validating it in similar conditions. Another proposal is presented in~\cite{mmwave_eurico}, when \ac{ac-rlnc} techniques are used to improve the UDP protocol, to support \ac{llc} and \ac{urllc} services under the lossy links.


\subsection*{Use-case 6: ICN deployment}

The \ac{icn} paradigm is considered to be one of the most promising candidates to overcome the drawbacks of host-centric architectures whose main flaws are in large-scale mobile distributed wireless environments such as \ac{iot} scenarios due to node mobility, dynamic topologies and intermittent connectivity~\cite{Djama:cc:2020}.
Despite the touted virtues of the \ac{icn} paradigm, and its several implementations, such as the \ac{ndn}, it is rare to find implementations or testbeds where the paradigm can be tested under the same conditions as networks based on Internet protocols. Even recent works such as \cite{Abane:fgcs:2019,Papadopoulos:CICN:2021,Yang:CICN:2021} discuss experiments in controlled environments. Furthermore, as far as we know, there is no \ac{icn} available on an extensive wireless communication infrastructure installed in a smart city. 

\begin{figure}[t]
	\centering
	\includegraphics[width=0.95\linewidth]{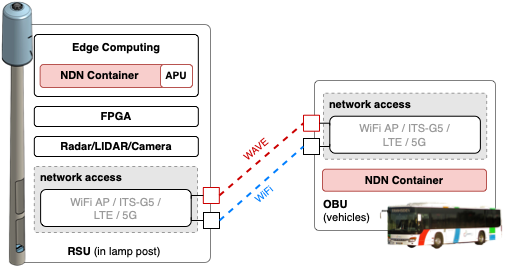} 
	\caption{NDN over ATCLL infrastructure.}
	\label{fig:ndn-over-steam}
\end{figure}

In this way, we have made a deployment of an \ac{ndn} solution, designed for mobile IoT environments~\cite{bib:Gameiro20,bib:Hernandez22}, over the \ac{atcll} communication infrastructure and assessed its performance in a content dissemination scenario. Two \ac{ndn} containers were installed on two network elements of the ATCLL, in a \ac{rsu} and in an \acp{obu}, as illustrated in Figure~\ref{fig:ndn-over-steam}.

To demonstrate the testbed flexibility in hosting multiple protocols, we compared the transmission process, regarding the transmission delay and number of NDN packets, when sending a 6MB file (fragmented in 1403 \ac{ndn} chunks) using ITS-G5, WiFi, and both technologies simultaneously, always using the \ac{ndn} paradigm.

\begin{table}[t]
\caption{Dissemination of 6MB file using NDN in the ATCLL infrastructure.}
\label{tab:ndn-over-steam}
\begin{center}
\begin{tabular}{ c||c|c|c }
\toprule
Tech & Time (s) & outInts & inData \\
\midrule
ITS-G5 & 25 & 1403 & 1403 \\  
WiFi & 82 &  1403 & 1403 \\ 
Both & 42 (14/42) & 1422 (711/711) & 1422 (711/711)\\
\bottomrule
\end{tabular}
\end{center}
\end{table}

Table \ref{tab:ndn-over-steam} shows that the simultaneous transmission by ITS-G5 (14s) and WiFi (42s) has its performance marked by the slower one. However, the option of the two technologies opens up a range of new heuristics to be applied in the \ac{ndn} forwarder module, the one responsible to select the outgoing interface of data chunks.


\section{Conclusions}
\label{sec:conclusions}

The large majority of existing contributions related to connectivity in different scenarios, including mobile scenarios with people, bicycles, and vehicles still rely on numerical computations and computer simulations, not addressing the impairments of real environments. 

In this article we present a living lab platform that allows the support of research activities in a real environment with real users, making them a beneficial part of the research. ATCLL is a facility deployed as a city-scale testbed with a large number of Internet-of-Things devices, with communication, sensing and computing capabilities, built on fiber and mmWave links, integrating a communication network with radio terminals, multiprotocol, spread throughout the city. The ATCLL supports a wide range of services and applications: \ac{iot}, intelligent transportation systems and assisted driving, environmental monitoring, emergency and safety, and more. 

The data of the platform is gathered through edge computing and a datacenter where the services, data platform, apps, 5G core and network functions are deployed in virtual machines running in the cloud. The presented results demonstrate a dashboard with
visibility of real-time data and examples of data collected (traffic radar, camera and WiFi sniffing). Performance results were also plotted, demonstrating the coverage of the communication technologies, the usage of edge computing and the implementation of the \ac{sdn}-based vehicular network. Finally, several use cases were tested in the infrastructure, in particular mobility characterization, collision avoidance with \acp{vru}, emergency vehicles communication, a data platform for industrial internet-of-things to cloud continuum, mmWave backhaul research and an \ac{icn} deployment. 

This facility and platform will be further improved and enhanced by federating it, a process which is already taking part as current work. 

\section*{Acknowledgements}
We would like to thank André Mourato and Tomás Freitas for the involvement in the development of the mobility services; Ricardo Torres and João Milheiro for the deployment of the spectral probes and SDR units; Alexandre Figueiredo and Tânia Ferreira for the research activities described in Use-case 5; and Luis Gameiro and Gustavo Amaral for the experiments described in Use-case 6.

This work has been funded by the European Regional Development Fund – ERDF, included in the “Urban Innovative Actions” programme, through project Aveiro STEAM City (UIA03-084).

\bibliographystyle{IEEEtran}
\bibliography{main}
\end{document}